\documentclass[preprintnumbers,nofootinbib,noshowpacs,eqsecnum,prd,superscriptaddress,letterpaper]{revtex4}

\usepackage{graphicx}
\usepackage{amsmath,amssymb}
\usepackage{url}
\usepackage{multirow}
\usepackage{hyperref,color}
\graphicspath{{figs/}}
\usepackage{xcolor}

\newcommand{\sla}[1]{/\!\!\!\!#1}

\newcommand{\Dfb}{\mbox{$\raisebox{2mm}{\boldmath ${}^\leftrightarrow$}
\hspace{-4mm} D$}}
\newcommand{\Dfba}{\mbox{$\raisebox{2mm}{\boldmath ${}^\leftrightarrow$}\hspace{-4mm} D^a$}}

\newcommand {\fbw}     {f_{BW}}
\newcommand {\fpone}   {f_{\Phi,1}}
\newcommand {\fwww}    {f_{WWW}}
\newcommand {\fw}      {f_{W}}
\newcommand {\fb}      {f_{B}}
\newcommand {\fww}     {f_{WW}}
\newcommand {\fbb}     {f_{BB}}
\newcommand {\fgg}     {f_{GG}}
\newcommand {\fptwo}   {f_{\Phi,2}}

\newcommand {\obw}     {{\cal O}_{BW}}
\newcommand {\opone}   {{\cal O}_{\Phi,1}}
\newcommand {\owww}    {{\cal O}_{WWW}}
\newcommand {\ow}      {{\cal O}_{W}}
\newcommand {\ob}      {{\cal O}_{B}}
\newcommand {\oww}     {{\cal O}_{WW}}
\newcommand {\obb}     {{\cal O}_{BB}}
\newcommand {\ogg}     {{\cal O}_{GG}}
\newcommand {\optwo}   {{\cal O}_{\Phi,2}}

\newcommand {\oqthree}  {{\cal O}^{(3)}_{\Phi,Q}}
\newcommand {\oqone}  {{\cal O}^{(1)}_{\Phi,Q}}
\newcommand {\our}  {{\cal O}^{(1)}_{\Phi,u}}
\newcommand {\odr}  {{\cal O}^{(1)}_{\Phi,d}}
\newcommand {\oud}  {{\cal O}^{(1)}_{\Phi,ud}}
\newcommand {\oer}  {{\cal O}^{(3)}_{\Phi,e}}
\newcommand {\ollll}  {{\cal O}_{LLLL}}
\newcommand {\ot}  {{\cal O}_{u\Phi,33}}
\newcommand {\obo}  {{\cal O}_{d\Phi,33}}
\newcommand {\ota}  {{\cal O}_{e\Phi,33}}
\newcommand {\omu}  {{\cal O}_{e\Phi,22}}            
\newcommand {\fqthree}  {f^{(3)}_{\Phi,Q}}
\newcommand {\fqone}  {f^{(1)}_{\Phi,Q}}
\newcommand {\fur}  {f^{(1)}_{\Phi,u}}
\newcommand {\fdr}  {f^{(1)}_{\Phi,d}}
\newcommand {\fud}  {f^{(1)}_{\Phi,ud}}
\newcommand {\fer}  {f^{(1)}_{\Phi,e}}
\newcommand {\fllll}  {f_{LLLL}}
\newcommand {\fbo}  {f_{b}}
\newcommand {\fta}  {f_{\tau}}
\newcommand {\ft}  {f_{t}}
\newcommand {\fm}  {f_{\mu}}
\preprint{\bf YITP-SB-18-39} 

\begin{document}

\title{Electroweak Sector Under Scrutiny: A Combined Analysis of LHC
and Electroweak Precision Data}
\author{ Eduardo da Silva Almeida}
\email{eduardo.silva.almeida@usp.br}
\affiliation{Instituto de F\'isica, Universidade de S\~ao Paulo, S\~ao Paulo, Brazil}
\author{Alexandre Alves}
\email{aalves.unifesp@gmail.com}
\affiliation{Departamento de F\'isica, Universidade Federal de S\~ao Paulo, UNIFESP,
Diadema, S\~ao Paulo, Brazil}
\author{N. Rosa-Agostinho}
\email{nuno.f.agostinho@gmail.com}
\affiliation{Departament  de  Fisica  Quantica  i  Astrofisica
 and  Institut  de  Ciencies  del  Cosmos,  Universitat
 de Barcelona, Diagonal 647, E-08028 Barcelona, Spain}
\author{Oscar J. P. \'Eboli}
\email{oeboli@gmail.com}
\affiliation{Instituto de F\'isica, Universidade de S\~ao Paulo,
 S\~ao Paulo, Brazil}
\author{M.~C.~Gonzalez--Garcia}
\email{maria.gonzalez-garcia@stonybrook.edu}
\affiliation{Departament  de  Fisica  Quantica  i  Astrofisica
 and  Institut  de  Ciencies  del  Cosmos,  Universitat
 de Barcelona, Diagonal 647, E-08028 Barcelona, Spain}
\affiliation{Instituci\'o Catalana de Recerca i Estudis Avancats (ICREA)
Pg. Lluis  Companys  23,  08010 Barcelona, Spain.}
\affiliation{C.N. Yang Institute for Theoretical Physics, Stony Brook University, Stony Brook NY11794-3849,  USA}
%

\begin{abstract}
  We perform a comprehensive study of the Higgs couplings, gauge-boson
  couplings to fermions and triple gauge boson vertices. We work in
  the framework of effective theories including the effects of the
  dimension-six operators contributing to these observables. We
  determine the presently allowed range for the coefficients of these
  operators via a 20 parameter global fit to the electroweak precision
  data, as well as electroweak diboson and Higgs production data from
  LHC Run 1 and 2. We quantify the improvement on the determination of
  the 20 Wilson coefficients by the inclusion of the Run 2 results.
  In particular we present a novel analysis of the ATLAS    Run 2
  36.1 $\rm fb^{-1}$  data on
  the transverse mass distribution of $W^+W^-$ and $W^\pm Z$ in the
  leptonic channel which allow for stronger tests of the triple gauge
  boson vertices.  We discuss the discrete (quasi)-degeneracies
  existing in the parameter space of operator coefficients relevant
  for the Higgs couplings to fermions and gauge bosons. In particular
  we show how the inclusion of the incipient $tH$ data can break those
  degeneracies in the determination of the top-Higgs coupling. We also
  discuss and quantify the effect of keeping the terms quadratic in
  the Wilson coefficients in the analysis and we show the
    importance of the Higgs data to constrain some of the operators
    that modify the triple gauge boson couplings in the linear
    regime.

\end{abstract}

\maketitle

\section{Introduction}
\label{sec:intro}

Since the discovery of a new state resembling the SM Higgs
boson~\cite{ Aad:2012tfa, Chatrchyan:2012xdj}, the CERN Large Hadron
Collider (LHC) has accumulated an impressive amount of data that allow
new searches for extensions of the Standard Model (SM), as well as
detailed studies of the SM predictions. Presently, the negative
results of the search for direct new physics effects indicate that new
states are probably heavy, therefore, there might exist a mass gap
between the SM states and the new ones. In this scenario, hints on the new
physics can manifest itself first through deviations from the SM
predictions.  \smallskip

Against this background, effective lagrangians raise as a most
adequate tool for scientific advancement.  The effective lagrangian
approach~\cite{Weinberg:1978kz,Georgi:1985kw,Donoghue:1992dd} is a
model--independent framework, which, using as inputs the low energy
particle contents and symmetries, describes new physics that is
expected to manifest itself directly at an energy scale $\Lambda$
larger than the scale at which the experiments are performed, by
including in the lagrangian higher--dimension operators.  In this
context, and within the present experimental results, we proceed by
assuming that the observed scalar {\rm belongs indeed to} a light
electroweak doublet and that the $SU(2)_L \otimes U(1)_Y$
symmetry is linearly realized in the effective theory.  Barring
effects associated with violation of total lepton number, the lowest
order operators which can be built are of dimension six.  The
coefficients of these dimension--six operators parametrize our
ignorance of the new physics effects and our task at hand is to
determine them using the available data.\smallskip

In this work we follow this road by performing a comprehensive
analysis of the observables related to the electroweak sector, which
at present allows for precision tests of the couplings between
electroweak gauge bosons and fermions, triple electroweak gauge
couplings (TGC) and the couplings of the Higgs to fermions and gauge
bosons. The first two sets of couplings allow us to probe the
$SU(2)_L \otimes U(1)_Y$ gauge structure of the SM, while the last one
aims the study of the couplings of the recently discovered scalar,
{\em i.e.}  to directly probe the electroweak symmetry breaking
mechanism. In this respect, this work extends and updates our previous
partial constrains on Higgs anomalous couplings and
TGC's~\cite{Corbett:2012dm, Corbett:2012ja, Corbett:2015ksa,
  Butter:2016cvz}. \smallskip

In the framework of the effective lagrangian described above (also
lately labeled in the literature as the SMEFT) our study involves
twenty dimension--six operators whose coefficients are determined by
means of a global fit to the relevant data, including low energy
electroweak precision measurements as well as LHC data on gauge boson
pair production and Higgs observables. The global analysis approach
(see Ref.~\cite{Ellis:2018gqa, Grojean:2018dqj} for recent related
works) is becoming mandatory because within the present LHC
statistics, changes in the couplings of gauge bosons to fermions, even
within the constraints of electroweak precision data (EWPD) can lead
to modifications of the kinematical distributions in gauge boson pair
production at LHC of comparable size to the ones stemming from the
purely anomalous
TGC~\cite{Zhang:2016zsp,Baglio:2017bfe,Alves:2018nof}.  And these, in
turn, influence the determination of the Wilson coefficients for the
operators entering the Higgs observables.  \smallskip

With this aim we briefly introduce in Sec.~\ref{sec:thframe} the set
of dimension--six operators included in our study and point out the
main sources of (quasi)-degenerate solutions which appear in the
analysis. Section~\ref{sec:frame} contains a brief description of the
data and statistical treatment applied to it. For the Higgs
observables this includes the final results of the LHC Run 1 and the
most up to date from Run 2 in terms of signal strengths or ratios of
cross sections and branching ratios. Concerning the gauge boson pair
production data, besides the final results of the LHC Run 1, we
perform a novel analysis of the ATLAS data on transverse mass
distribution of $W^+W^-$ and $W^\pm Z$ in the leptonic channel. The
body of ours results are discussed in Sec.~\ref{sec:results}, which
we present in terms of the different sectors tested:
gauge-boson-fermion couplings, TGC's and Higgs couplings.  One
particularly interesting observation is the relevance of the incipient
$tH$ data to break possible degeneracies in the determination of the
top Yukawa coupling.  We finish Sec.~\ref{sec:results} with a
quantification of the relevance of keeping the terms quadratic in the
Wilson coefficients in the analysis.  In that respect our results show
that the accumulated statistics on the Higgs observables is starting
to be large enough for meaningful constraints on the Wilson
coefficients of some of the operators to be imposed from their
interference with the SM contributions.  We summarize our findings in
Sec.~\ref{sec:summary}. \smallskip

\section{Theoretical Framework}
\label{sec:thframe}

Up to now there is no direct evidence of new states being produced at
the LHC. Therefore, we parametrize possible deviations from the SM by
higher-dimension operators:
\begin{equation}
   {\cal L}_{\rm eff} = {\cal L}_{\rm SM} + \sum_{n>4,j}
   \frac{f_{n,j}}{\Lambda^{n-4}} {\cal O}_{n,j} \;,
\label{eq:gen}
\end{equation}
where the SM $SU(3)_C \otimes SU(2)_L \otimes U(1)_Y$ gauge symmetry
is realized linearly in the ${\cal O}_{n,j}$ operators. For the sake
of simplicity we retain only the dimension--six operators that conserve
$C$, $P$ and lepton and baryon numbers.  The first higher-order
operator is of dimension five~\cite{Weinberg:1979sa}, however, it does
not contribute to the LHC physics due to the strong constraints
originating from the neutrino sector. The first operators to play a
significant role at the LHC are of dimension six, {\em i.e.}
$n=6$. It is well known that there are 59 independent dimension--six
operators~\cite{Grzadkowski:2010es}, up to flavor and hermitian
conjugation. Using the freedom in the choice of basis~\cite{
  Politzer:1980me, Georgi:1991ch, Arzt:1993gz, Simma:1993ky} due to
the use of equations of motion (EOM), we work in that of Hagiwara,
Ishihara, Szalapski, and Zeppenfeld (HISZ)~\cite{Hagiwara:1993ck,
  Hagiwara:1996kf} for the pure bosonic operators. \smallskip

In what follows we focus on the subset of the dimension--six operator
basis that impact the precision electroweak data
(EWPD)~\cite{ALEPH:2005ab}, TGC's, and Higgs physics.  The EWPD
observables receive linear contributions from seven operators
involving fermions, gauge bosons and the Higgs field:
\begin{equation}
  \begin{array}{l@{\hspace{1cm}}l@{\hspace{1cm}}l}
& 
   {\cal O}^{(1)}_{\Phi L,ij} =\Phi^\dagger (i\, \Dfb_\mu \Phi) 
(\bar L_{i}\gamma^\mu L_{j}) \;\;,
& 
{\cal O}^{(3)}_{\Phi L,ij}
=\Phi^\dagger (i\,{\Dfba}_{\!\!\mu} \Phi) 
(\bar L_{i}\gamma^\mu T_a L_{j}) \;\;, 
\\
&
&
\\
& 
{\cal O}^{(1)}_{\Phi Q,ij}=
\Phi^\dagger (i\,\Dfb_\mu \Phi)  
(\bar Q_i\gamma^\mu Q_{j}) \;\;,
& 
{\cal O}^{(3)}_{\Phi Q,ij} =\Phi^\dagger (i\,{\Dfba}_{\!\!\mu} \Phi) 
(\bar Q_i\gamma^\mu T_a Q_j) \;\;,
\\
&
&
\\
& {\cal O}^{(1)}_{\Phi u,ij}=\Phi^\dagger (i\,\Dfb_\mu \Phi) 
(\bar u_{R_i}\gamma^\mu u_{R_j}) \;\;,
& {\cal O}^{(1)}_{\Phi d,ij}=\Phi^\dagger (i\,\Dfb_\mu \Phi) 
(\bar d_{R_i}\gamma^\mu d_{R_j}) \;\;,
\\
&
&
\\
& 
{\cal O}^{(1)}_{\Phi e,ij}=\Phi^\dagger (i\Dfb_\mu \Phi) 
(\bar e_{R_i}\gamma^\mu e_{R_j})  \;\;, &
\label{eq:ewpd-op1}
\end{array}
\end{equation}
together with a purely four fermion operator:
\begin{equation}
  {\cal O}_{LLLL}
  =(\bar L \gamma^\mu L)(\bar L \gamma^\mu L) \;\;.
\label{eq:ewpd-op2}
\end{equation}
In addition to the above fermionic operators there are two bosonic
operators that contribute to the EWPD
\begin{equation}
\mathcal{O}_{BW}		
= \Phi^\dagger\widehat{B}_{\mu\nu}\widehat{W}^{\mu\nu}\Phi  
\;\;\;\;\hbox{ and }\;\;\;\;
  \mathcal{O}_{\Phi,1}		
=	(D_\mu\Phi)^\dagger\Phi\Phi^\dagger(D^\mu\Phi) \;\;.
\label{eq:bw-phi1}
\end{equation}
In the equations above $\Phi$ stands for the SM Higgs doublet while
lepton (quark) doublet is denoted by $L$ ($Q$) and $f_R$ are the
$SU(2)_L$ singlet fermions, where $i, j$ are family indices. In
addition, we defined
$\Phi^\dagger\Dfb_\mu\Phi= \Phi^\dagger D_\mu\Phi-(D_\mu\Phi)^\dagger
\Phi$ and
$\Phi^\dagger \Dfba_{\!\!\mu} \Phi= \Phi^\dagger T^a D_\mu
\Phi-(D_\mu\Phi)^\dagger T^a \Phi$ with $T^a=\sigma^a/2$. Also, we
defined $\widehat{B}_{\mu\nu} \equiv i(g^\prime/2)B_{\mu\nu}$ and
$\widehat{W}_{\mu\nu} \equiv i(g/2)\sigma^aW^a_{\mu\nu}$, with $g$ and
$g^\prime$ being the $SU(2)_L$ and $U(1)_Y$ gauge couplings
respectively. The Pauli matrices are denoted by $\sigma^a$.
\smallskip

The seven operators in Eq.~\eqref{eq:ewpd-op1} modify directly the
coupling of the $Z$ to fermion pairs while
${\cal O}^{(3)}_{\Phi Q,ij}$ contributes also to the $W$ couplings to
left quarks. ${\cal O}_{LLLL}$'s, ${\cal O}_{BW}$ and
${\cal O}_{\Phi,1}$ contributions are ubiquitous and stem from their
effect on the finite renormalization of the SM fields and couplings
once the lagrangian is canonically normalized.  In particular
${\cal O}_{LLLL}$ gives a finite contribution to the Fermi constant
while the operators ${\cal O}_{BW}$, and ${\cal O}_{\Phi,1}$ lead to
finite correction to the $S$ and $T$ oblique parameters respectively.
Furthermore, in the analysis of EWPD we did not considered six dipole
operators whose interference with the SM contributions vanish for
EWPD observables. \smallskip

In what follows, in order to avoid the existence of blind
directions~\cite{DeRujula:1991ufe, Elias-Miro:2013mua} in the EWPD
analyses, we used the freedom associated to the use of EOM to remove
the operator combinations
\begin{equation}
  \sum_i {\cal O}^{(1)}_{\Phi L,ii} \;\;\;{\rm and}\;\;\;
  \sum_i {\cal O}^{(3)}_{\Phi L,ii} \;\;
\label{eq:EOMred}  
\end{equation}
%
from our operator basis.  Moreover, we assume no family mixing in the
above operators to prevent the generation of too large flavor
violation and for simplicity we consider the operators to be
generation independent; hereafter, we drop the generation indexes for
these operators.  With these hypothesis, the operators
${\cal O}^{(1)}_{\Phi L}$ and ${\cal O}^{(3)}_{\Phi L}$ are removed by
the use of EOM~\cite{Corbett:2012ja}.  \smallskip

With the above assumptions, in our basis, only the operator
${\cal O}^{(1)}_{\Phi e}$ modifies the $Z$ coupling to leptons,
while there are additional contributions to the $Z$-quark pair
vertices originating from $ {\cal O}^{(1)}_{\Phi u}$,
$ {\cal O}^{(1)}_{\Phi d}$, ${\cal O}^{(1)}_{\Phi Q}$, and
${\cal O}^{(3)}_{\Phi Q}$.  Moreover, the $W$ coupling to fermions
receives extra contributions from ${\cal O}^{(3)}_{\Phi Q}$; see
Table~\ref{tab:coupl2}. \smallskip

\begin{table}
{\footnotesize  \begin{tabular}{|c|c|c|c|c|c|c|c|c|c|}
\hline 
& $H\bar{f}f$ 
& $Z\bar{q}{q}$
& $Z\bar{l}{l}$   & $W\bar{u}{d}$   & $W\bar{l}{\nu}$   
\\ \hline 
$\mathcal{O}_{BW}$  & & X & X & X & {X}
\\ \hline
${\mathcal{O}_{\Phi,1}}$  & X 
& {X} & {X} & {X} &{X} 
\\ \hline
$\mathcal{O}_{\Phi,2}$  & X
& &  & &
\\ \hline
${\cal O}^{(1)}_{\Phi Q}$,
${\cal O}^{(1)}_{\Phi u}$,
${\cal O}^{(1)}_{\Phi d}$ 
&
& X  &  & &
 \\ \hline 
${\cal O}^{(3)}_{\Phi Q}$,
 &
 & X  &  & X &
\\ \hline 
${\cal O}^{(1)}_{\Phi e}$,
&
&   & X & &
\\ \hline 
${\cal O}^{(1)}_{\Phi ud}$
&
&   &  & X &
\\\hline
${\cal O}_{u\Phi,33}$
& X (if $f=t$)
&   &  &  &
\\\hline
${\cal O}_{d\Phi,33}$
& X (if $f=b$)
&   &  &  &
\\\hline
${\cal O}_{e\Phi,33}$
& X (if $f=\tau$)
&   &  &  &
\\\hline
${\cal O}_{u\Phi,22}$
& X (if $f=\mu$)
&   &  &  &
\\\hline
\end{tabular}
\caption{Anomalous couplings to fermions generated by the
  dimension--six operators considered in the analysis.}
\label{tab:coupl2}
}
\end{table}

Altogether the part of the dimension--six effective lagrangian that
contributes to the EWPD is
\begin{eqnarray}
 \Delta  {\cal L}_{\rm eff}^{\rm EWPD}
  &=& \frac{f^{(1)}_{\Phi Q}}{\Lambda^2}   {\cal O}^{(1)}_{\Phi Q}
  +     \frac{f^{(3)}_{\Phi Q}}{\Lambda^2}   {\cal O}^{(3)}_{\Phi Q}
  +     \frac{f^{(1)}_{\Phi u}}{\Lambda^2}   {\cal O}^{(1)}_{\Phi u}
  +     \frac{f^{(1)}_{\Phi d}}{\Lambda^2}   {\cal O}^{(1)}_{\Phi d}
  +     \frac{f^{(1)}_{\Phi e}}{\Lambda^2}   {\cal O}^{(1)}_{\Phi e}
\nonumber
\\
&+&
\frac{f_{BW}}{\Lambda^2} {\cal O}_{BW}
+ \frac{f_{\Phi,1}}{\Lambda^2} {\cal O}_{\Phi,1}
  +     \frac{f_{LLLL}}{\Lambda^2}   {\cal O}_{LLLL} \;\;. 
\label{eq:leff-ewpd}
\end{eqnarray}
As is well known, the above operators are strongly constrained by the
EWPD observables~\cite{Corbett:2017qgl}. \smallskip


Data on electroweak diboson production processes (here on EWDBD) at the LHC,
$pp\rightarrow W^+W^-$ and $pp\rightarrow Z W^\pm$, can be used to
study operators that change the $W$ and $Z$ couplings to fermions, as
well as TGC. These processes receive contributions from the previously
discussed operators, as well as from the fermionic operator
\begin{equation}
{\cal O}^{(1)}_{\Phi ud}=\tilde\Phi^\dagger (i\,\Dfb_\mu \Phi) 
(\bar u_{R}\gamma^\mu d_{R} +{\rm h.c.}) \;\;,
\end{equation}
that modifies the couplings of $W$'s to right-handed quark pairs and does not interfere
with the SM contributions to the EWPD observables at the order considered
in the analysis.  In addition, TGC's are also modified by  
two additional dimension--six operators that include Higgs and
electroweak gauge fields in the HISZ basis 
\begin{equation}
\begin{array}{lll}
 \mathcal{O}_{W}		
=	(D_\mu\Phi)^\dagger\widehat{W}^{\mu\nu}(D_\nu\Phi)  
& \hbox{ and }  
&\mathcal{O}_{B}		
=	(D_\mu\Phi)^\dagger\widehat{B}^{\mu\nu}(D_\nu\Phi)  \;\;,
\end{array}
\label{eq:b-w}
\end{equation}
and by one operator that contains exclusively gauge bosons
\begin{equation}
  \mathcal{O}_{WWW} = {\rm Tr}[\widehat{W}_{\mu}^{\nu}
    \widehat{W}_{\nu}^{\rho}\widehat{W}_{\rho}^{\mu}] \;\;.
\label{eq:www}
\end{equation}
We present in Table~\ref{tab:coupl3} the anomalous TGC stemming from
the dimension--six operators in our basis.  It is interesting to notice
that dimension--six operators do not give rise to anomalous TGC among
neutral gauge bosons. We define the effective lagrangian of the
operators that contribute to TGC in addition to the ones participating
in the EWPD analysis as
\begin{equation}
\Delta{\cal L}_{\rm eff}^{\rm TGC} = 
 \frac{f_{WWW}}{\Lambda^2} {\cal  O}_{WWW}
+ \frac{f_{W}}{\Lambda^2} {\cal O}_{W}
+ \frac{f_{B}}{\Lambda^2} {\cal O}_{B}
+ \frac{f^{(1)}_{\Phi ud}}{\Lambda^2} {\cal O}^{(1)}_{\Phi ud} \;\;.
\label{eq:leff-tgc}
\end{equation}

Concerning Higgs processes, in order to quantify possible deviations
from the SM predictions we must consider some additional
dimension--six operators that contain the Higgs field $\Phi$. In particular the
following fermionic operators modify the Yukawa couplings of the
Higgs boson
\begin{equation}
{\cal O}_{e\Phi,ij}=(\Phi^\dagger\Phi)(\bar L_i \Phi e_{R,j}) 
\;\;\;,\;\;\;
{\cal O}_{u\Phi,ij}=(\Phi^\dagger\Phi)(\bar Q_i \tilde \Phi u_{R,j})
\;\;\;\hbox{and}\;\;\;
{\cal O}_{d\Phi,ij}=(\Phi^\dagger\Phi)(\bar Q_i \Phi d_{R,j}) \;\; ,
\label{eq:hffop}
\end{equation}
with $\tilde \Phi=i\sigma_2\Phi^*$ and $i,j$ being generation
indices. In this work we consider only the diagonal couplings of Higgs
to the third family and also to muon pairs (which are the only ones
being currently tested at LHC) -- that is, we work with the effective
lagrangian
\begin{equation}
\Delta   {\cal L}_{\rm eff}^{\rm Yuk} = 
\frac{f_\mu m_\mu}{\Lambda^2 v} {\cal O}_{e\Phi,22} 
+ \frac{f_\tau m_\tau}{ \Lambda^2 v} {\cal O}_{e\Phi,33} 
+ \frac{f_b m_b}{ \Lambda^2 v} {\cal O}_{d\Phi,33} 
+ \frac{f_t m_t}{ \Lambda^2 v} {\cal O}_{u\Phi,33} 
+ \text{ h.c.}
\label{eq:leff-yuk}
\end{equation}
Finally the couplings of the Higgs to gauge bosons are further modified
by the bosonic operators 
\begin{equation}
\begin{array}{lll}
 {\cal O}_{GG}  = \Phi^\dagger \Phi \; G_{\mu\nu}^a G^{a\mu\nu}  
& \;\;,
& {\cal O}_{WW}  = \Phi^{\dagger} \hat{W}_{\mu \nu} \hat{W}^{\mu \nu} \Phi  \;\;,
\\
&
&
\\
 {\cal O}_{BB}  = \Phi^{\dagger} \hat{B}_{\mu \nu} \hat{B}^{\mu \nu} \Phi 
&\;\;,
& {\cal O}_{\Phi,2} = \frac{1}{2} \partial^\mu\left ( \Phi^\dagger \Phi \right)
                            \partial_\mu\left ( \Phi^\dagger \Phi
  \right ) \;\;.
\label{eq:higgs}  
\end{array}
\end{equation}
The operator ${\cal O}_{\Phi,2}$ modifies all Higgs couplings due to a
finite renormalization of the Higgs wave function.
We keep its effects explicitly in the modification of the Higgs
vertices generated from the dimension--four part of the Lagrangian,
${\cal  L}_{\rm SM}$. Corrections to vertices generated by other
dimension--six operators
are absorbed in the redefinition of the corresponding Wilson
coefficient. 

As for  the effective couplings to gluons, in what follows we work in the
convention that the Wilson coefficient of the operator
$ {\cal O}_{GG}$ contains not only contributions of new possible
colored states appearing in the loop connecting gluon pairs to the
Higgs, but also the effects of the anomalous operators that modify the
SM Yukawa coupling to bottom and top quarks running in the loop. The
effective lagrangian associated to these operators is
\begin{equation}
\Delta{\cal L}_{\rm eff}^{\rm HVV} = 
- \frac{\alpha_s }{8 \pi} \frac{f_{GG}}{\Lambda^2} {\cal O}_{GG}  
+ \frac{f_{BB}}{\Lambda^2} {\cal O}_{BB} 
+ \frac{f_{WW}}{\Lambda^2} {\cal O}_{WW} 
+ \frac{f_{\Phi,2}}{\Lambda^2} {\cal O}_{\Phi,2} \;\;.
\label{eq:leff-higgs}
\end{equation}
Notice that we rescale the Wilson coefficient $f_{GG}/\Lambda^2$ of
the gluon-gluon operator to include a loop suppression factor such
that its limits are of the same order of the Wilson coefficients of
other operators. \smallskip

\begin{table}
{\footnotesize  \begin{tabular}{|c|c|c|c|c|c|c|}
\hline 
& $ZWW$ & $\gamma WW$ & $H \gamma \gamma$ & $HZZ$ & $HZ\gamma$ &$HWW$ 

\\ \hline 
$\mathcal{O}_{WWW}$  
& X & X& & & & 
\\ \hline 
$\mathcal{O}_{W}$ & X & X & & X & X & X 
\\ \hline 
$\mathcal{O}_{B}$ & X & X &  & X & X & 
\\ \hline 
$\mathcal{O}_{BW}$ & X & X & X & X & X & X 
\\ \hline 
$\mathcal{O}_{WW}$ 
&  & & X & X & X& X 
\\ \hline
$\mathcal{O}_{BB}$ & &  & X & X & X & 
\\ \hline
${\mathcal{O}_{\Phi,1}}$ & X & & & X & & X 
\\ \hline
$\mathcal{O}_{\Phi,2}$ & & & & X & & X 
\\ \hline
\end{tabular}
\caption{Anomalous couplings of gauge and Higgs bosons induced the
  dimension--six operators the we consider.}
\label{tab:coupl3}
}
\end{table}


In summary, the total effective lagrangian that we consider is
\begin{equation}
{\cal L}_{\rm eff} = {\cal L}_{\rm SM} 
+\Delta {\cal L}_{\rm eff}^{\rm EWPD}
+ \Delta {\cal L}_{\rm eff}^{\rm TGC}
+ \Delta {\cal L}_{\rm eff}^{\rm Yuk}
+ \Delta {\cal L}_{\rm eff}^{\rm HVV} \;\;.
\label{eq:leff}
\end{equation}
Tables~\ref{tab:coupl2} and \ref{tab:coupl3} show the anomalous three
point vertices generated by the effective lagrangian in
Eq.~(\ref{eq:leff}); the explicit form of the couplings and the
different Lorentz structures generated can be found, for example, in
Refs.~\cite{Corbett:2012ja, Corbett:2017qgl, Corbett:2014ora,
  Alves:2018nof} to which we refer the reader for details. \smallskip

In brief, EWPD constrains at tree level only the eight Wilson
coefficients of the operators appearing in
$ \Delta{\cal L}_{\rm eff}^{\rm EWPD}$, while the TGC analysis of
EWDBD receives contributions from the twelve operators in
$\Delta{\cal L}_{\rm eff}^{\rm EWPD}$ and
$\Delta{\cal L}_{\rm eff}^{\rm TGC}$. On the other hand, nineteen of
the twenty operators in ${\cal L}_{\rm eff}$ are required to study the
different Higgs production and decay channels at LHC; in fact, only
${\cal O}_{WWW}$ does not contribute to Higgs processes. Altogether,
at dimension six, the global analysis of EWPD, EWDBD and Higgs processes
involves the 20 operators in Eq.~\eqref{eq:leff}. \smallskip

As we will see in the following section, at present there is enough
experimental information to individually bound the 20 Wilson coefficients
but there are still important discrete (quasi-)degeneracies. They can
be understood in terms of sign flips of the couplings of the Higgs
to gauge bosons and to fermions with respect to the SM. 
For instance, the coefficient of the $H W^+_\mu W^{- \mu}$ vertex is
\begin{equation}
\left ( \frac{g^2 v}{2} \right ) 
\left [
1 - \frac{v^2}{4} 
\left (
 \frac{f_{\Phi, 1}}{\Lambda^2}+ 2 \frac{f_{\Phi,2}}{\Lambda^2}
\right ) 
\right ] \;.
\label{eq:vert-hww} 
\end{equation}
Since $f_{\Phi,1}/\Lambda^2$ possesses a stringent bound from EWPD, we
anticipate a degeneracy with the SM results for $f_{\Phi,2}/\Lambda^2=0$
and around $f_{\Phi,2}/\Lambda^2=4 / v^2 \sim 65$ TeV$^{-2}$.
These points in parameter space are also nearly degenerate for the vertex
$H Z_\mu Z^\mu$. \smallskip

As for the Higgs couplings to fermions, anomalous interactions
can also lead to Yukawa couplings of the order
of the SM ones but with a different sign as the coefficient of the $H
\bar{f} f$ vertex is now
\begin{equation}
- \frac{m_f}{v} \left [
1 - \frac{v^2}{2} 
\left (
\frac{f_{\Phi,2}}{\Lambda^2} + \sqrt{2} \frac{f_f}{\Lambda^2} 
\right )
\right ]
\label{eq:vert-yuk}
\end{equation}
where $f= \mu, \tau, b, t$. Since $f_{\Phi,2}/\Lambda^2$ has two
different values compatible with flipping the sign of the SM $HVV$ coupling, we
can anticipate that $f_f/\Lambda^2$ will have $2\times 2$ degenerate
SM-like solutions, two corresponding to $f_f/\Lambda=0$, and the other two with 
$f_f/\Lambda= \pm 2 \sqrt{2}/v^2\sim 45$ TeV$^{-2}$. \smallskip

A further source of degeneracy is the effective gluon-gluon-Higgs
interaction associated to the operator $H G_{\mu\nu}^a G^{a,\mu\nu}$
whose coefficient is
\begin{equation}
-\frac{1}{4} G^{gg}_{\rm SM} - \frac{\alpha_S v}{8 \pi}\frac{f_{GG}}{\Lambda^2} \;\;,
\label{eq:vert-gluglu}
\end{equation}
where $G^{gg}_{\rm SM}\sim -5.3\times 10^{-2}$ TeV$^{-1}$ summarizes the SM
one--loop contribution. Flipping the sign of the SM contribution leads to
the existence of a SM-like solution for
$f_{GG}/\Lambda^2\sim -4\pi/(v\alpha_s) G_{\rm gg, SM}\sim 25$ TeV$^{-2}$.
The equivalent effect is present in the photon-photon-Higgs coupling 
$H F_{\mu\nu} F^{\mu\nu}$ with a coefficient
\begin{equation}
  -\frac{1}{4} G^{\gamma\gamma}_{\rm SM} +
  \frac{e^2 v}{4} \frac{f_{WW}+f_{BB}-f_{BW}}{\Lambda^2} \;\;,
\label{eq:vert-gaga}
\end{equation}
where $G^{\gamma\gamma}_{\rm SM}\sim 3.3\times 10^{-2}$ TeV$^{-1}$,
and a SM-like solutions for the $H\gamma\gamma$ decay can be found for
$(f_{WW}+f_{BB}-f_{BW})/\Lambda^2 \sim 2/(v \, e^2) G_{\rm
  \gamma\gamma,SM} \sim 3$ TeV$^{-2}$.  This degeneracy, however, is
only approximate because EWPD independently constrains $\fbw$ and the
measurement of the effective photon-Z-coupling
$H F_{\mu\nu} Z^{\mu\nu}$ bounds a different combination of $f_{WW}$,
$f_{BB}$ and $f_{BW}$. \smallskip

\section{ANALYSES FRAMEWORK}
\label{sec:frame}


In order to constrain the Wilson coefficients of the dimension--six
operators in the effective lagrangian Eq.~(\ref{eq:leff}), we
considered the EWPD, diboson production and Higgs signal strengths.
In the EWPD analysis we take into account 15 observables of which 12
are $Z$ observables~\cite{ALEPH:2005ab}:
\begin{equation*}
\Gamma_Z \;\;,\;\;
\sigma_{h}^{0} \;\;,\;\;
{\cal A}_{\ell}(\tau^{\rm pol}) \;\;,\;\;
R^0_\ell \;\;,\;\;
{\cal A}_{\ell}({\rm SLD}) \;\;,\;\;
A_{\rm FB}^{0,l} \;\;,\;\;
R^0_c \;\;,\;\;
 R^0_b \;\;,\;\;
{\cal  A}_{c} \;\;,\;\;
 {\cal A}_{b} \;\;,\;\;
A_{\rm FB}^{0,c}\;\;,\;\;
\hbox{ and} \;\;
A_{\rm FB}^{0,b}  \hbox{ (SLD/LEP-I)}\;\;\; ,
\end{equation*}
supplemented by three $W$ observables
\begin{equation*}
  M_W   \;\;,\;\; \Gamma_W \;\;\hbox{  and}\;\; \hbox{Br}( W\to {\ell\nu})
\end{equation*}
that are, respectively, its average mass taken
from~\cite{Olive:2016xmw}, its width from
LEP2/Tevatron~\cite{ALEPH:2010aa}, and the leptonic $W$ branching
ratio for which the average in Ref.~\cite{Olive:2016xmw} is
considered. In order to perform the statistical analysis we
constructed a $\chi^2$ function for the EWPD
\begin{equation}
\chi^2_{\rm EWPD}(\fbw,\fpone,\fqthree,\fqone,\fur,\fdr,\fer,\fllll) \;.
\label{eq:chi2ewpd}
\end{equation}
We include in our EWPD analysis the correlations among the above
observables as displayed in Ref.~\cite{ALEPH:2005ab}. Furthermore, the
SM predictions and their uncertainties due to variations of the SM
parameters were extracted from~\cite{Ciuchini:2014dea}.  For further
details of this part of the statistical analysis we refer the reader
to Refs.~\cite{Corbett:2017qgl, Alves:2018nof} .  \smallskip


The structure of the electroweak triple gauge boson coupling has been
the subject of direct scrutiny in gauge boson pair production at
LEP2~\cite{lep2} and the Run 1 of LHC where the ATLAS and CMS
collaborations have used their full data samples on
$W^+ W^-$~\cite{Aad:2016wpd,Khachatryan:2015sga} and
$W^\pm Z$~\cite{Aad:2016ett, Khachatryan:2016poo} productions to
constrain the possible deviations of TGC's from the SM structure in
terms of the effective Lorentz invariant parametrization of
Ref.~\cite{Hagiwara:1986vm} or in terms of coefficients of some of the
relevant dimension--six operators. For Run 2, the number of
experimental studies aiming at deriving the corresponding limits is
still rather sparse. In particular ATLAS~\cite{ATLAS:2016qzn} has
presented some results on bounds on TGC couplings from $WZ$ production
but still with data collected with 13.3 fb$^{-1}$. With this limited
luminosity, the TGC sensitivity is still below that of Run 1. However,
ATLAS has also presented results on their measurements of diboson
production at 13 TeV with higher luminosity and this data can be used
to set better constraints on TGC. With this aim in here we use the
ATLAS results on $WZ$ production~\cite{ATLAS:2018ogj} and on
$WW$~\cite{Aaboud:2017gsl} both with 36.1 fb$^{-1}$ as we describe
next. \smallskip

In order to obtain the bounds on the Wilson coefficients in the
effective lagrangian in Eq.~\eqref{eq:leff} we study the $W^+W^-$ and
$W^\pm Z$ productions in the leptonic channel using the available
kinematic distribution that is most sensitive for TGC analysis. More
specifically, the channels that we analyze and their kinematical
distributions are:
\begin{center}
\begin{tabular}{ l|lcll}
\hline 
Channel ($a$) & Distribution & \# bins   &\hspace*{0.2cm} Data set & \hspace*{0.2cm}Int Lum 
\\ [0mm]
\hline
$WW\rightarrow \ell^+\ell^{\prime -}+\sla{E}_T\; (0j)$
& $p^{\rm leading, lepton}_{T}$
& 3 & ATLAS 8 TeV, &20.3 fb$^{-1}$~\cite{Aad:2016wpd}
\\[0mm]
$WW\rightarrow \ell^+\ell^{(\prime) -}+\sla{E}_T\; (0j)$
& $m_{\ell\ell^{(\prime)}}$ & 8 & CMS 8 TeV, &19.4 fb$^{-1}$~\cite{Khachatryan:2015sga}
\\[0mm]
$WZ\rightarrow \ell^+\ell^{-}\ell^{(\prime)\pm}$ & $m_{T}^{WZ}$ & 6 & ATLAS 8 TeV, & 20.3 fb$^{-1}$~\cite{Aad:2016ett}
\\[0mm]
$WZ\rightarrow \ell^+\ell^{-}\ell^{(\prime)\pm}+\sla{E}_T$ & $Z$ candidate $p_{T}^{\ell\ell}$ & 10 & CMS 8 TeV, &19.6 fb$^{-1}$~\cite{Khachatryan:2016poo}
\\[0mm]
$WW\rightarrow e^\pm \mu^\mp+\sla{E}_T\; (0j)$
&  $m_T$ & 17 & 
ATLAS 13 TeV, &36.1 fb$^{-1}$~\cite{Aaboud:2017gsl}
\\[0mm]
$WZ\rightarrow \ell^+\ell^{-}\ell^{(\prime)\pm}$ 
&  $m_{T}^{WZ}$ & 6 
& ATLAS 13 TeV, &36.1 fb$^{-1}$~\cite{ATLAS:2018ogj}
\\[0mm]
\hline
\end{tabular}
\end{center}
For each experiment and channel, we extract from the experimental
publications the observed event rates in each bin, $N^{a}_{i,\rm d}$,
as well as the background expectations $N^{a}_{i,\rm bck}$, and the SM
$W^+W^-$ ($W^\pm Z$) predictions, $N^{a}_{i,\rm sm}$. \smallskip

For details of the analysis of EWDBD from Run 1 we refer the reader to
Ref.~\cite{Alves:2018nof} that contains our procedure, as well as its
validation against the TGC results from the experimental
collaborations.  \smallskip

Concerning the Run 2 EWDBD analysis, for the $W^+W^-$ final state we
study the transverse mass distribution in the ATLAS 13 TeV ggF sample.
We extract from Fig.~4 in Ref.~\cite{Aaboud:2017gsl} both the data,
the non-WW backgrounds, as well as the SM $WW$ contributions in each
of the 17 bins in the transverse mass variable
\begin{equation}
  m_T \equiv \sqrt{
(E_T^{\ell\ell} + E_T^{\rm miss})^2 - | \vec{p}_T^{\ell\ell} +
\vec{p}_T^{\rm miss} |^2
}
\end{equation}
with
$E_T^{{}\ \ell\ell} = \sqrt{| \vec{p}_T^{\ell\ell}|^2 + m^2_{\ell\ell}
}$ and the transverse momentum (invariant mass) of the lepton pair
denoted by $ \vec{p}_T^{\ell\ell}$ ($m_{\ell\ell}$).  The statistical
uncertainty is given by $\sqrt{N^{WW}_{i,\rm d}}$ where we combine the
contents of the last 3 bins to ensure gaussianity.  Theoretical and
systematic uncertainties can be found in Tables~5-7 of the same
reference. For the 13 TeV $W^\pm Z$ final state, we make the analysis
using the transverse mass distribution in Fig.~4c in
Ref.~\cite{ATLAS:2018ogj} which is presented in terms of $W^\pm Z$
signal events in 6 bins covering all values of $m^{WZ}_T$ (last bin
containing all data above 600 GeV) and already background subtracted.
From the lower panel of the same figure we read the statistical, total
experimental and theoretical uncertainties in each bin\footnote{Since
  the data points give the number of $WZ$ signal events after
  background subtraction their statistical error read from the figure
  are much larger than $\sqrt{N^{WZ}_{i,\rm d}}$.}. \smallskip

With that information the procedure to obtain the relevant kinematical
distributions predicted in presence of the dimension--six operators is
the same followed for our Run 1 EWDBD analysis~\cite{Alves:2018nof}.
First we simulate the $W^+W^-$ and $W^\pm Z$ productions using
\textsc{MadGraph5}~\cite{Alwall:2014hca} with the UFO files for our
effective lagrangian generated with
\textsc{FeynRules}~\cite{Christensen:2008py, Alloul:2013bka}.  We
employ \textsc{PYTHIA6.4}~\cite{Sjostrand:2006za} to perform the
parton shower, while the fast detector simulation is carried out with
\textsc{Delphes}~\cite{deFavereau:2013fsa}. In order to account for
higher order corrections and additional detector effects we simulate
SM $W^+W^-$ and $W^\pm Z$ productions in the fiducial region requiring
the same cuts and isolation criteria adopted by the corresponding
ATLAS studies, and normalize our results bin by bin to the
experimental collaboration predictions for the kinematical
distributions under consideration.  Then we apply these correction
factors to our simulated $WV$ distributions in the presence of the
anomalous couplings.  \smallskip

The statistical confrontation of these predictions with the LHC Run 2
data is made by means of a binned log-likelihood function based on the
contents of the different bins in the kinematical distribution of each
channel. Besides the statistical errors we incorporate the systematic
and theoretical uncertainties added in quadrature to the uncorrelated
statistical error in each bin assuming some partial correlation among
them which we estimate to range between 30\% and 70\% with the
information provided. With this, we build the corresponding
$\chi^2_{\rm EWDBD}$ which we combine with the EWPD bounds so we have
\begin{equation}
\chi^2_{\rm EWPD+EWDBD}(\fb, \fw, \fwww, \fbw, \fpone, \fqthree, \fqone,
\fur, \fdr, \fud, \fer,\fllll) \;.
\label{eq:chi2ewtgc}
\end{equation}

As for Higgs processes, we use the available data
from  Runs 1 and 2 from the following sources
\begin{center} 
\begin{tabular}{ lcc }
\hline 
Source     & Int.Luminosity (fb$^{-1}$) &\hspace*{0.3cm} \# Data points
\\
\hline
ATLAS+CMS at 7 \& 8 TeV~\cite{Khachatryan:2016vau} [Table 8, Fig 27]
&  5 \& 20 & 20+1
\\
ATLAS at 13 TeV~\cite{ATLAS:2018doi} [Figs. 6,7] & 79.8 & 9  
\\
CMS at 13 TeV~\cite{Sirunyan:2018koj} [Table 3] & 35.9 & 24 
\\
\hline
ATLAS at 8 TeV~\cite{Aad:2015gba} ($\gamma Z$) & 20 & 1 
\\
ATLAS at 13 TeV~\cite{Aaboud:2017uhw} ($\gamma Z$) & 36.1 & 1 
\\
ATLAS at 13 TeV~\cite{ATLAS-CONF-2018-026} ($\mu^+\mu^-$)  & 36.1 & 1 
\\
\hline
\end{tabular}
\end{center}
that provide us signal strengths or ratios of cross sections and
branching ratios. The first three references above contain information
on all production mechanisms and almost all decay channels in the
Figures and Tables given in the first column. Moreover, these
references also provide the correlation matrix among the observables,
as well as statistic and systematic errors (for
Ref.~\cite{Sirunyan:2018koj} the correlation matrix can be found
in~\cite{cmstwiki}).  The fourth and fifth references contains
information on the rare decay mode $\gamma Z$ while the last one on
the $\mu^+ \mu^-$ channel.  \smallskip

The statistical comparison of our effective theory predictions with
the LHC Runs 1 and 2 data is made by means of a $\chi^2_{\rm Higgs}$
function based on these 22 (Run 1) + 35 (Run 2) data points.  Adding
this to the analysis of EWPD and EWDBD we construct our global 20
dimensional statistical function
\begin{equation}
  \chi^2_{\rm EWPD+EWDBD+Higgs} (\fb, \fw, \fwww, \fbb,\fww,\fbw,\fgg, \fpone,
  \fptwo,\fqthree, \fqone, \fur, \fdr, \fud, \fer,\fllll,\fbo,\ft,\fta,\fm) \;.
\end{equation}

\section{Results}
\label{sec:results}

We present in Figs.~\ref{fig:fer_r2}--\ref{fig:yuk_r2} $\Delta\chi^2$
profiles (in all cases marginalized with respect to all undisplayed 
parameters involved in the analysis) for the Wilson coefficients for three sets
of analyses which differ in the data samples included:
\begin{itemize}
\item EWPD: $\Delta\chi^2_{\rm EWPD}$ which constrains the 8
  coefficients in $\Delta{\cal L}_{\rm eff}^{\rm EWPD}$,
  Eq.~\eqref{eq:leff-ewpd}.  They are given by the green lines in
  Figs.~\ref{fig:fer_r2} and~\ref{fig:bos_r2}. This analysis is
  performed taking into account only the contributions to the
  observables that are linear in the anomalous Wilson coefficients;
  for further detail see Ref.~\cite{Corbett:2017qgl}.
\item EWPD+EWDBD: $\Delta\chi^2_{\rm EWPD+EWDBD}$ which limits the 12
  coefficients in
  $ \Delta {\cal L}_{\rm eff}^{\rm EWPD}+\Delta {\cal L}_{\rm
    eff}^{\rm TGC}$, Eqs.~\eqref{eq:leff-ewpd}
  and~\eqref{eq:leff-tgc}.  The results are depicted in
  Fig.~\ref{fig:tgv_r2}. In the evaluation of the predictions for
  EWDBD we have kept the contribution of the Wilson coefficients up to
  the quadratic order.
\item GLOBAL$\equiv$ EWPD+EWDBD+HIGGS: $\Delta\chi^2_{\rm EWPD+EWDBD+HIGGS}$
  which constrains the 20 coefficients in ${\cal L}_{\rm eff}$ in
  Eq.~\eqref{eq:leff} (
  (Eqs.~\eqref{eq:leff-ewpd}--Eqs.\eqref{eq:leff-higgs}).  They are
  the red, black, and dashed blue curves in
  Figs.~\ref{fig:fer_r2},~\ref{fig:bos_r2} and~\ref{fig:yuk_r2}.
  In the evaluation of the predictions for EWDBD and the Higgs data
  we have kept the contribution of the Wilson coefficients to the
  physical observables up to the quadratic order. 
\end{itemize}

\subsection{Gauge boson couplings to fermions}
\label{sec:ferm}

\begin{figure}[h!]
\centering
\includegraphics[width=0.85\textwidth]{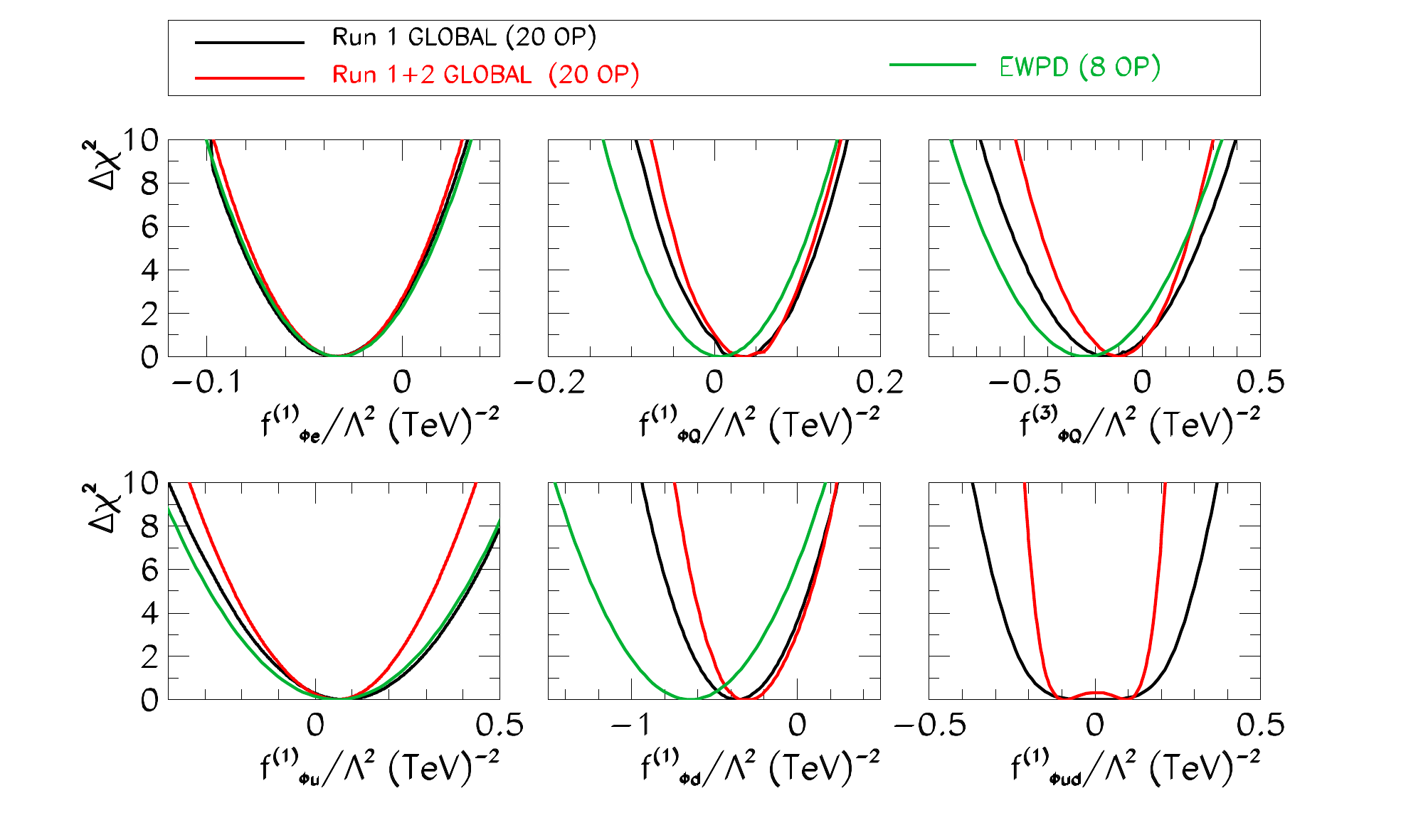}
\caption{$\Delta \chi^2$ as a function of the fermionic Wilson
  coefficients $\fer/\Lambda^2$, $\fqone/\Lambda^2$,
  $\fqthree/\Lambda^2$, $\fur/\Lambda^2$, $\fdr/\Lambda^2$, and
  $\fud/\Lambda^2$, as indicated in the panels after marginalizing
  over the remaining fit parameters. The green solid line stands for
  the fit of the EWPD that constrains only eight of twenty Wilson
  coefficients in Eq.~(\ref{eq:leff}). The black (red) solid line
  represents the twenty-parameter fit to the LHC Run 1 (and 2) data.}
  \label{fig:fer_r2}
\end{figure}

Our results concerning the determination of the Wilson coefficients
for the operators involving gauge boson and fermion fields and which
directly modify the gauge couplings to fermions are shown in
Fig.~\ref{fig:fer_r2}. As it is well known, EWPD yields strong bounds on
deviations of the SM predictions for the fermion-gauge interactions,
and this is quantified in the green curves in the figure. The
additional information provided by the inclusion of the LHC data from
EWDBD and Higgs observables (now in the larger 20 parameter space)
collected at Run 1 (and Run 2) are shown as the black (red)
curves. \smallskip

In the upper left panel of Fig.~\ref{fig:fer_r2}, we find the
$\Delta\chi^2$ dependence on $\fer/\Lambda^2$ which is the coefficient
of the only operator involving gauge couplings to leptons remaining in
the basis after applying the EOM.  This operator modifies the $Z$
coupling to right-handed leptons which were precisely tested at
LEP. On the contrary, at the LHC observables it enters only via its
contribution to the decay rate of the $Z$ boson to leptons in some of
the final states considered. Consequently, as seen in the figure, the
inclusion of the LHC data does not add any meaningful
information about this coefficient. \smallskip

The central and right upper panels in Fig.~\ref{fig:fer_r2} display
the $\Delta\chi^2$ dependence on the coefficients $\fqone/\Lambda^2$
and $\fqthree/\Lambda^2$, which correspond to operators modifying the
couplings of left-handed quarks to $Z$ and $W$ bosons. On the other
hand, the left and central lower panels correspond to the dependence
on $\fur/\Lambda^2$ and $\fdr/\Lambda^2$ which give corrections to the
$u_R$ and $d_R$ couplings to $Z$ respectively. Comparing the green
with the black and red lines we see that the impact of the inclusion
of the LHC results is still minor but not negligible, in particular
for $f^{(1)}_{\Phi d}/\Lambda^2$. The EWPD analysis favors
non-vanishing value for $f^{(1)}_{\Phi d}/\Lambda^2$ at 2$\sigma$, a
result driven by the 2.7$\sigma$ discrepancy between the observed
$A_{\rm FB}^{0,b}$ and the SM. On the contrary, no significant
discrepancy is observed between the relevant LHC observables, in
particular in EWDBD, and the SM predictions.  Hence there is a shift
towards zero of $f^{(1)}_{\Phi d}/\Lambda^2$ when including the LHC
data in the analysis. This slightly smaller tension results also into
the reduction of the globally allowed range. This behavior was
observed in Ref.~\cite{Alves:2018nof} for Run 1 data and the inclusion
of Run 2 results adds in this direction. \smallskip

Finally, in the right lower panel we show the $\Delta\chi^2$ on the
coefficient of $\oud$ operator. This operator induces a right-handed
coupling of the $W$ boson to quarks. At the linear level it does not
interfere with the SM and its effect has not been included in the EWPD
analysis. The dependence shown in the figure arises from its
contribution to the LHC observables which we keep up to the quadratic
order (notice that $\Delta \chi^2$ as a function of this coupling is
symmetric around zero even though its minimum is not exactly at zero).
The figure illustrates how including the effect of this
operator to that order leads to bounds on its coefficient which are
comparable to those of the other operators that modify the coupling of
electroweak gauge bosons to quarks and interfere with the
SM. \smallskip

\subsection{Triple anomalous gauge couplings in Diboson searches at Run 2}
\label{sec:tgc}

As mentioned in the previous section, for Run 2 the number of
experimental studies focused on deriving constraints in the size and
structure of TGC's is very limited and makes use only of a small
fraction of their collected data~\cite{ATLAS:2016qzn}. But the ATLAS
collaboration has presented results on diboson production in
Ref.~\cite{ATLAS:2018ogj,Aaboud:2017gsl} which we make use to test the
TGC's; see previous section for details on our construction of the
corresponding likelihood functions. \smallskip

\begin{figure}[h!]
\centering
\includegraphics[width=0.8\textwidth]{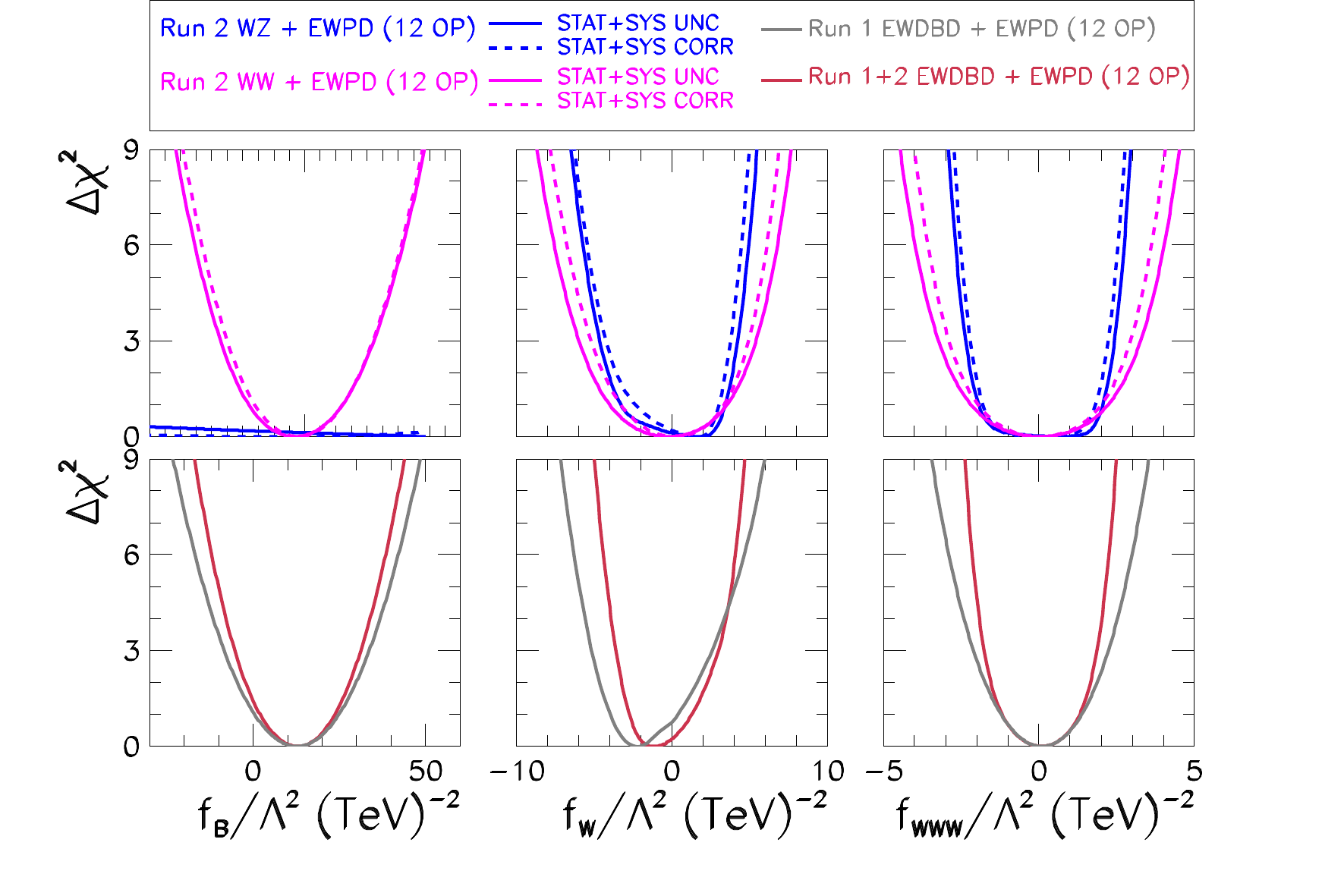}
\caption{$\Delta \chi^2$ dependence on the $f_B/\Lambda^2$ (left
  panel), $f_{W}/\Lambda^2$ (central panel) and $f_{WWW}/\Lambda^2$
  (right panel) parameters after the marginalization over the 11
  undisplayed fit parameters for the analysis of LHC EWDBD and EWPD.
  The upper panels show the results of our analysis of the ATLAS Run 2
  data on $WZ$ ~\cite{ATLAS:2018ogj} and on $WW$~\cite{Aaboud:2017gsl}
  transverse mass distributions. Full lines (dashed) correspond to
  assuming zero (total) correlation among the non-statistical
  uncertainties; see text for details.  The lower panels show the
  results of the analysis of the EWDBD from Run 1 of
  Ref.~\cite{Alves:2018nof} in combination with EWPD (black lines) and
  including also the results from ATLAS $WW$ and $WZ$ production at Run
  2 (red line).}
  \label{fig:tgv_r2}
\end{figure}

The results of our analysis of the ATLAS Run 2 $WW$ and $WZ$ leptonic
data~\cite{ATLAS:2018ogj,Aaboud:2017gsl} together with the EWPD in the
twelve-dimensional parameter space
\[
\left \{
\fb, \fw, \fwww, \fbw, \fpone, \fqthree, \fqone,
\fur, \fdr, \fud, \fer
\right \}
\]
are shown in the upper panels in Fig.~\ref{fig:tgv_r2}, where we plot
the one--dimensional $\Delta\chi^2$ distributions for the Wilson
coefficients of the ``TGC operators'' $\ob$, $\ow$ and $\owww$ after
marginalization over the 11 undisplayed coefficients.
As expected, the $WZ$ channel gives no constraint on $\ob$ while both
$WW$ and $WZ$ contribute with similar precision to the determination
of $\fw/\Lambda^2$ and $\fwww/\Lambda^2$. To illustrate the possible
effect of our assumptions on the correlations of the
systematic/theoretical uncertainties (labeled as SYS in the figure)
among the different bins we show the results obtained with full (zero)
correlation among those uncertainties in the dashed (solid) lines. As
seen in the figure the effect is small. \smallskip

In the lower panels we show the impact of adding the Run 2 $WW$ and
$WZ$ results to the analysis of the Run 1 diboson data of
Ref.~\cite{Alves:2018nof} which included data on $WW$ and $WZ$
channels from both ATLAS and CMS collected with $\sim$ 20 fb$^{-1}$ at
each experiment. 
Altogether we find that the combined ATLAS Run 2
diboson data constrains the operator coefficients with precision
similar (a bit better indeed) to that of the full Run 1 analysis. This
is expected from simple statistics of the integrated luminosity and
energy scaling. The combination of CMS and ATLAS Run 1 data accounts
for about 40 fb$^{-1}$ in each $WW$ and $WZ$ channels which is of the
order of the 36 fb$^{-1}$ of ATLAS Run 2 data.  Moreover, the total
cross section for diboson productions is about twice larger at Run 2
than at Run 1. \smallskip

\subsection{Higgs couplings}
\label{sec:higgs}

\begin{figure}[h!]
\centering
\includegraphics[width=0.85\textwidth]{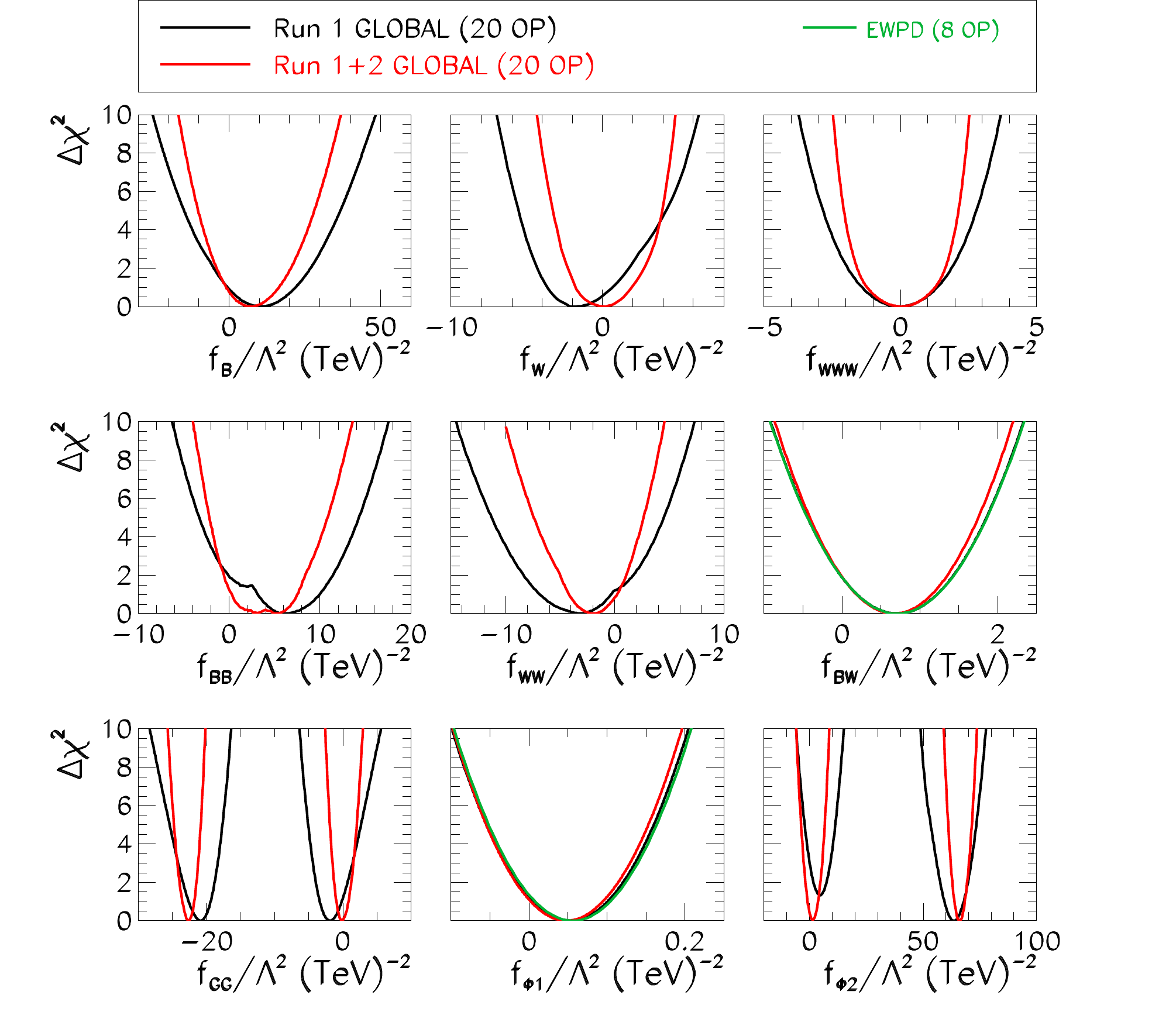}
\caption{$\Delta \chi^2$ dependence on the bosonic Wilson coefficients
  $f_{B}/\Lambda^2$, $f_{W}/\Lambda^2$, $f_{WWW}/\Lambda^2$,
  $f_{BB}/\Lambda^2$, $f_{WW}/\Lambda^2$, $f_{BW}/\Lambda^2$,
  $f_{GG}/\Lambda^2$, $f_{\Phi,1}/\Lambda^2$, and
  $f_{\Phi,2}/\Lambda^2$ as indicated in each panel.  The black (red)
  line stands for the results of the twenty-parameter fit using EWPD,
  EWDBD and Higgs data from LHC Run 1 (and 2). As before, the green
  line stands for the fit of only the EWPD. }
  \label{fig:bos_r2}
\end{figure}

\begin{figure}[h!]
\centering
\includegraphics[width=0.65\textwidth]{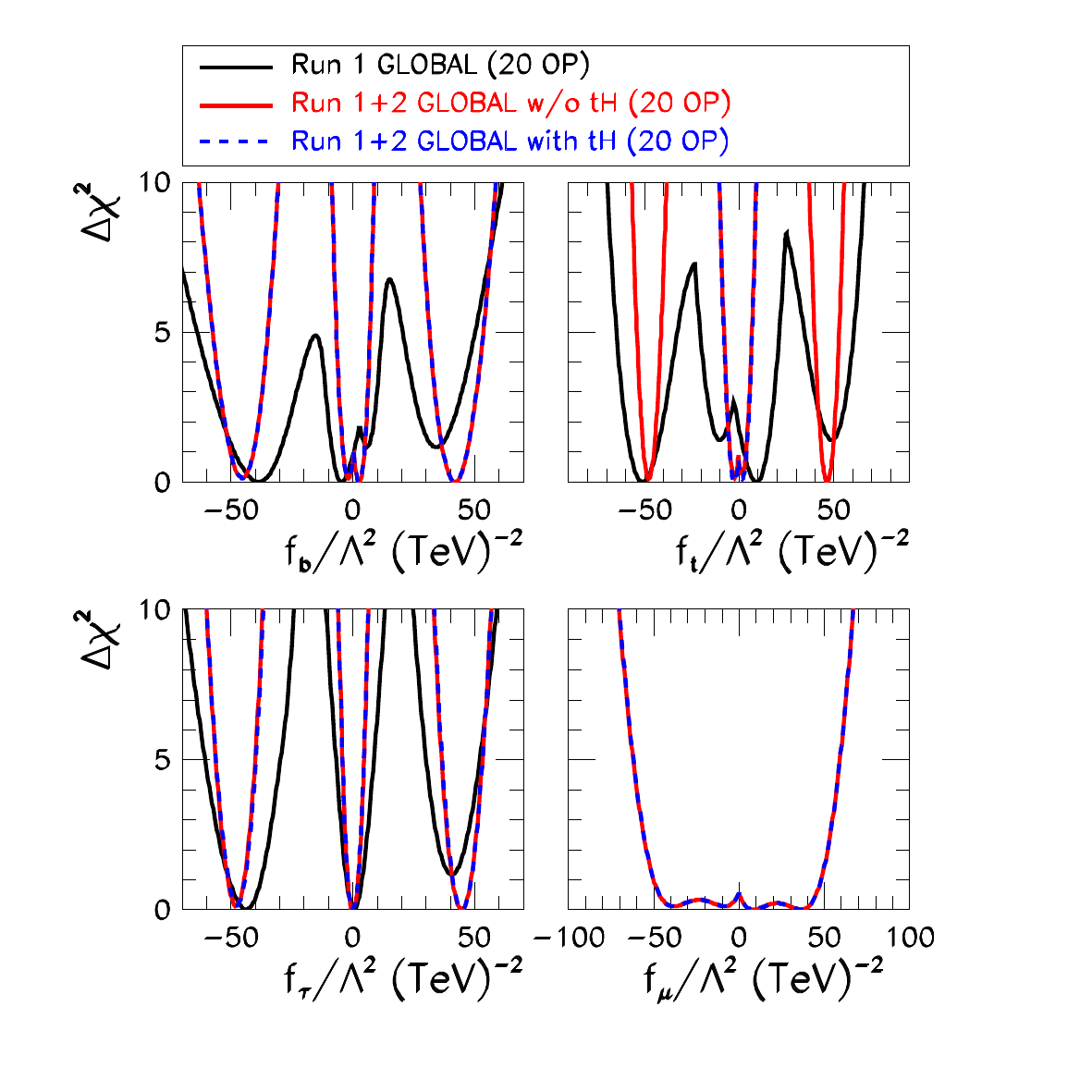}
\caption{$\Delta \chi^2$ dependence on the fermionic Wilson coefficients
  $f_{b}/\Lambda^2$, $f_{t}/\Lambda^2$, $f_{\tau }/\Lambda^2$, and
  $f_{\mu}/\Lambda^2$ as indicated in each panel. 
  The black (blue) line stands for the results
  of the twenty-parameter fit using EWPD, diboson production and Higgs
  data from LHC Run 1 (and 2).  The red line represents that results
  obtained using EWPD, diboson production and Higgs
  data from LHC Run 1 and 2 without the $tH$ contribution to the Higgs
top associate production cross section (see text for details).}
  \label{fig:yuk_r2}
\end{figure}

Our results concerning the determination of the Wilson coefficients
for the operators affecting the interactions of the Higgs field with
the gauge bosons and with fermions are shown in Figs.~\ref{fig:bos_r2}
and~\ref{fig:yuk_r2}, respectively.  In order to perform the most
general analysis of the Higgs boson couplings we use the full data set
that we presented in Section~\ref{sec:frame}, {\em i.e.} EWPD, EWDBD
and Higgs data, including the effect of the 20 operators in
Eq.~(\ref{eq:leff}). \smallskip

Figure~\ref{fig:bos_r2} depicts the one--dimensional $\Delta \chi^2$ as
a function of the Wilson coefficients of the pure bosonic operators in
Eq.~(\ref{eq:leff}) after fitting the EWPD and Run 1 (and 2) data on
Higgs and diboson productions. As expected, the most stringent
constraints are those on the oblique operators $\opone$ and $\obw$
that come from the EWPD with very little impact of the LHC
data~\cite{Alves:2018nof}. \smallskip

The first row of Fig.~\ref{fig:bos_r2} contains the $\Delta \chi^2$
distributions for the coefficients of the ``TGC operators''. Of those
only $\ob$ and $\ow$ enter both in TGC's and Higgs processes. For
completeness we include here the results of our global analysis on
$f_{WWW}/\Lambda^2$ also but we notice that $\owww$ does not involve
the Higgs field. As we can see, altogether the inclusion of the Run 2
data improves the bounds on the coefficients of the three ``TGC
operators'' by a factor ${\cal O}(25\%)$. Also comparing the results
for these operators to the second row of Fig.~\ref{fig:tgv_r2} we
learn that the inclusion of the Higgs data set strengthens the bounds
on $f_{B}/\Lambda^2$ and $f_{W}/\Lambda^2$ derived from the EWDBD
analysis by ${\cal O}(10$ --
$20\%)$~\cite{Corbett:2013pja}.\smallskip
%

\begin{figure}[h!]
\centering
\includegraphics[width=0.45\textwidth]{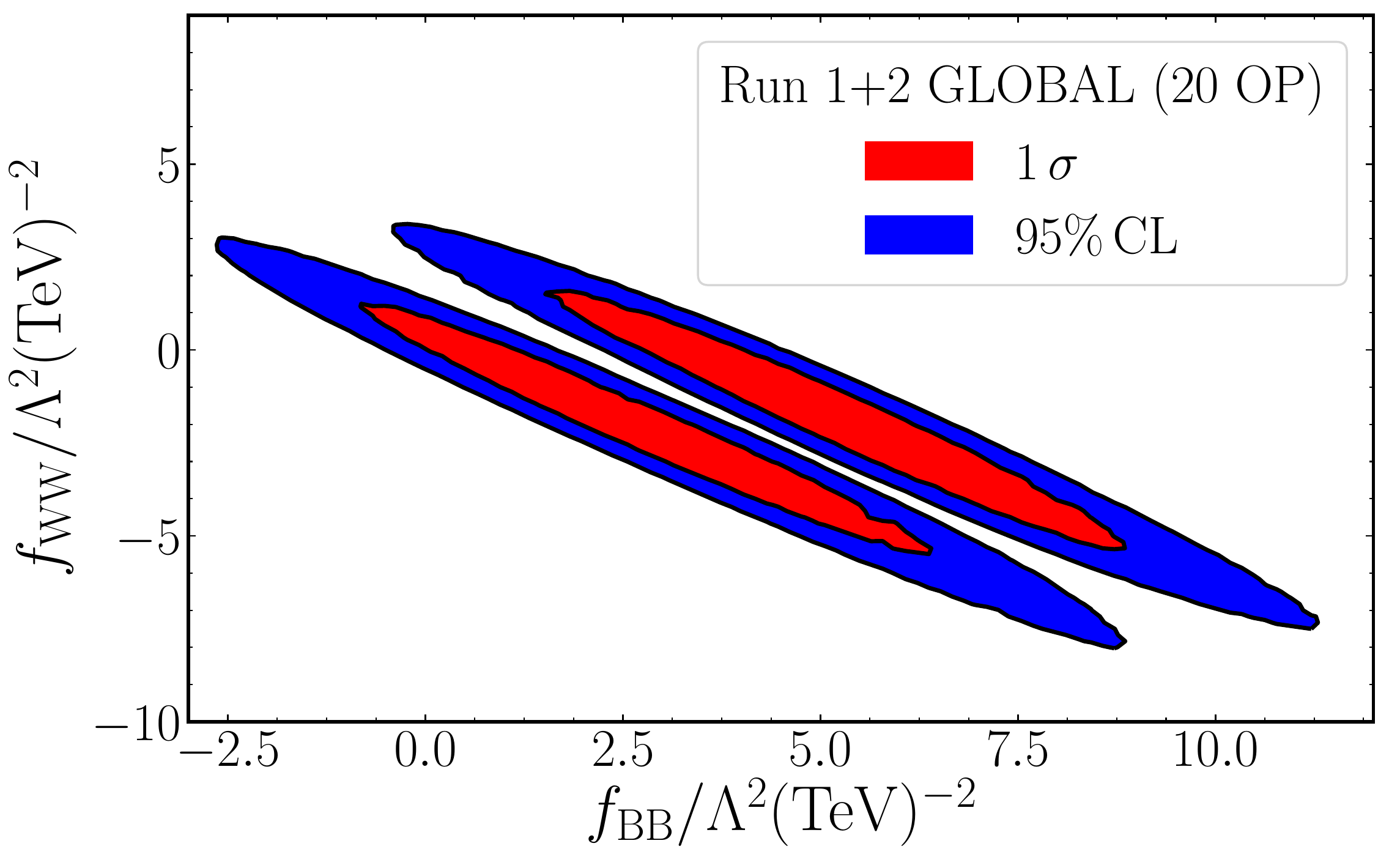}
\includegraphics[width=0.45\textwidth]{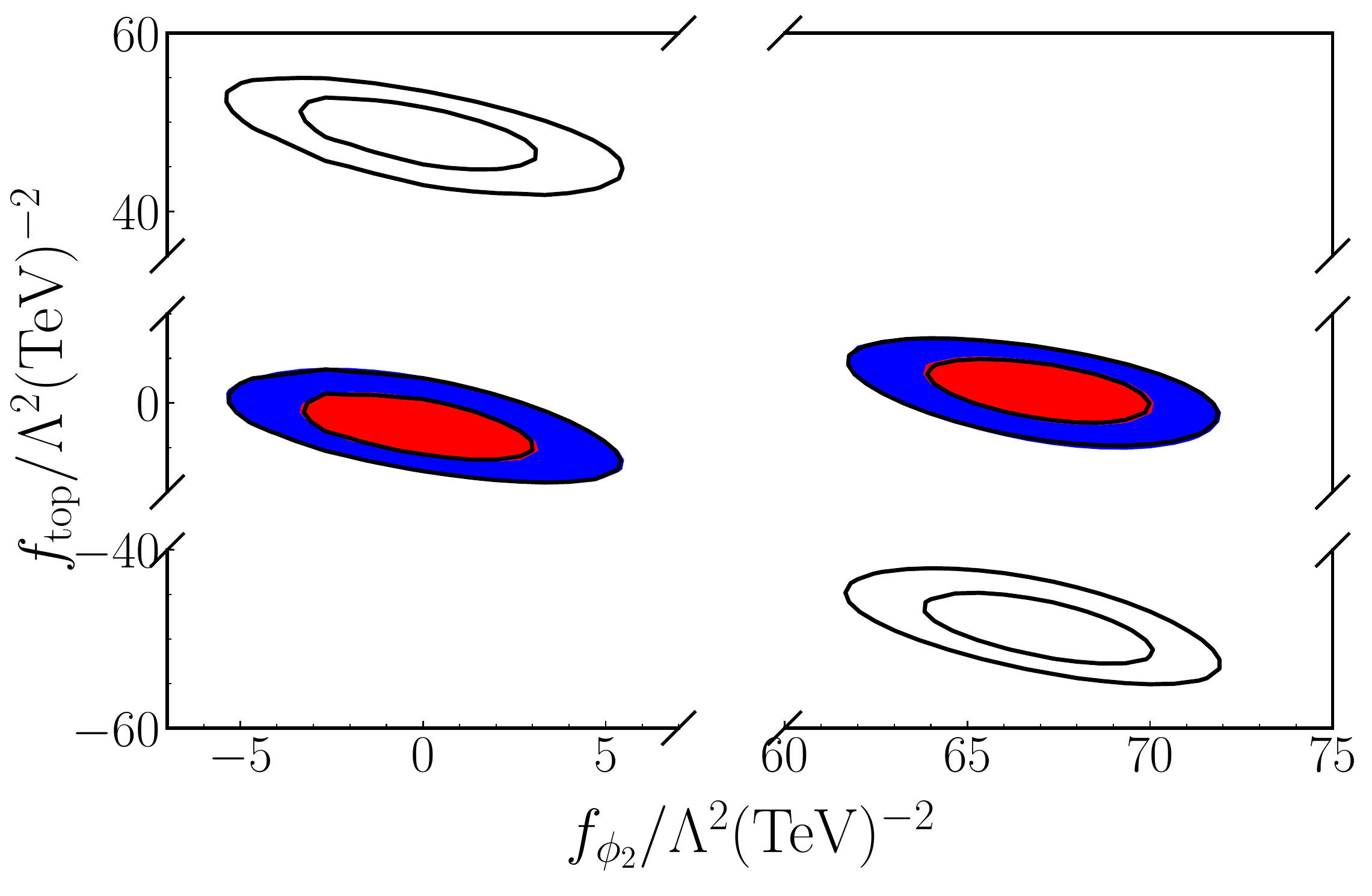}
\caption{$1\sigma$ and 95\% CL (2dof) allowed regions from the global analysis
  in the planes indicated in the axes. In the right panel the filled regions
  are obtained from the global analysis including the $tH$ contribution
  to the top Higgs associate production data of ATLAS at Run 2 while the
  void regions are the additional solutions allowed when the $tH$
  contribution is not included.}
\label{fig:correl}
\end{figure}

The operators $\obb$ and $\oww$ modify the Higgs decay into two
photons with a contribution proportional to
$f_{BB}/\Lambda^2 + f_{WW}/\Lambda^2$, therefore, introducing a strong
correlation between these operators~\cite{Corbett:2012dm} since the
decay rate for this Higgs channel is very well measured.  This is
illustrated in Fig.~\ref{fig:correl} where we show the two--dimensional
allowed regions from the global analysis for different pairs of Wilson
coefficients after marginalizing over the 18 undisplayed parameters in
each panel.  In particular on the left panel we show the degree of
anti-correlation between $f_{BB}/\Lambda^2$ and $f_{WW}/\Lambda^2$
still present after combination of Run 1 and 2 data. We also see the
two quasi-degenerate solutions discussed in Eq.\eqref{eq:vert-gaga},
the lower one containing the SM solution ($f_{WW}=f_{BB}=0$) and the
upper one with flipped sign of the $H\gamma\gamma$ coupling
``displaced'' by $\sim 3$ TeV$^{-2}$.  \smallskip

The lower row of Fig.~\ref{fig:bos_r2} contains the results for
$f_{GG}/\Lambda^2$ and $f_{\Phi,2}/\Lambda^2$ which clearly display
the discrete (quasi-) degeneracies explained in
Section~\ref{sec:thframe} (see discussions around
Eqs.~\eqref{eq:vert-gluglu} and \eqref{eq:vert-hww}) associated with
the reversion of the sign of the $Hgg$ and $HVV$ ($V=Z,W$) couplings
respectively. Notice that the two solutions for $f_{GG}/\Lambda^2$ are
completely equivalent in the analysis since this coupling only appears
in one vertex. On the other hand, the operator $\optwo$ modifies
universally all the SM-like Higgs couplings to gauge bosons and
fermions.  For each of them there are two degenerate solutions due to
the total reversal of the coupling sign, but they would only lie at
exactly the same values of $f_{\Phi,2}/\Lambda^2$ if all the couplings
were measured to have the same ratio to their SM value.  The
quasi-degeneracy in $f_{\Phi,2}/\Lambda^2$ present in our global
analysis originates from the lack of tension between the SM
predictions and the data for all processes so values around
$f_{\Phi,2}/\Lambda^2\sim 0$ and
$f_{\Phi,2}/\Lambda^2\sim \frac{4}{v^2}$ with all other couplings zero
can lead to a good global description of the data.  As seen in the
figure, the addition of Run 2 data has contributed in this direction.
\smallskip


We display in Fig.~\ref{fig:yuk_r2} the $\Delta \chi^2$ dependence
on the Wilson coefficients of the operators generating anomalous
Yukawa couplings. Comparing the results obtained using just the LHC
Run 1 data set (black curve) with the ones that contains the LHC Run 2
data (red curve), we can see that the Run 2 data is essential to
better constrain these Wilson coefficients. For instance, we can
witness the emergence of the three discrete solutions for
$f_b/\Lambda^2$ and $f_\tau/\Lambda^2$ which originate from
Eqs.~(\ref{eq:vert-hww}) and (\ref{eq:vert-yuk}) as explained in
Section~\ref{sec:thframe}. On the other hand, it is clear from this
figure that the data on Higgs decay into muon pairs is still incipient
and within the present precision the allowed regions around the three
minima merge into a unique allowed range. \smallskip

We show in this figure the results of the global analysis under two
assumptions for the top-Higgs associate production in Run 2. As
described in Ref.~\cite{ATLAS:2018doi} both $ttH$ and $tH$ (including
$tHW$ and $tHj$) contribute to the cross section ratio given in
Fig.~\ref{fig:linear} of that reference. But with the information
provided, it is not possible to determine the relative contribution of
$tH$ vs $ttH$ to the reconstructed total cross section ratio.  We show
the results for two extreme assumptions: a ratio of the $tH$ to $ttH$
contribution as predicted by the model ({\em i.e.} exactly same
reconstruction efficiency for both subprocesses), shown as the dashed
blue lines in the figure, and a negligible small contribution from
$tH$ shown in the red line.  For consistency, we see that the results
for all non-top Yukawa couplings are exactly the same for the two
analyses. For $f_t/\Lambda^2$ we find that including a ``full'' $tH$
contribution results into the total breaking of the degeneracies and
eliminates solutions other than the ones around $f_t/\Lambda^2=0$.
This can also be seen in the right panel in Fig.~\ref{fig:correl}
where we show the allowed regions in the plane $f_t/\Lambda^2$ vs
$f_{\Phi,2}/\Lambda^2$. The void and colored regions of this panel
show the four solutions resulting from Eqs.~(\ref{eq:vert-hww}) and
(\ref{eq:vert-yuk}) explained in Section~\ref{sec:thframe} which are
quasi-degenerated as long as no information on the sign of the $ttH$
coupling is available. On the other hand, the colored regions are the
only ones allowed once the information on $tH$ is included in the
analysis as described above.
This is expected as the $tH$ scattering amplitude receives
  contributions from the $ttH$ and $VVH$ vertices, therefore, being
  sensitive to the relative sign of the different diagrams
  contributing to it. In fact, the sign with respect to the SM of the
  vertices $ttH$ and $VVH$ are the same in the surviving colored
  regions in Fig.~\ref{fig:correl}. 
This result clearly illustrates the importance of the measurement of the $tH$
production rate to unambiguously determine the coupling of the Higgs to
the top quark. \smallskip

\subsection{Results at linear order}
\label{sec:linear}

In order to interpret Eq.~(\ref{eq:gen}) as a systematic expansion in
the large mass parameter $\Lambda$~\cite{Weinberg:1978kz}, its
contributions to observables at the lowest order ($1/\Lambda$) are
given the interference between anomalous and SM terms. In principle, if one 
keeps the quadratic contributions on the Wilson coefficients of
dimension--six operators one should include the interference of
dimension--eight operators with the SM as they are of the same order.
\smallskip

Up to this point we consider the effective lagrangian
Eq.~(\ref{eq:leff}) as a straw man that we use to probe the standard
model couplings assuming that it contains all information on possible
new physics. The results obtained are thus physically meaningful as
long as no large cancellations between the dimension--six quadratic
terms and the (here absent) linear dimension--eight SM interference are
expected.  Furthermore, this is a pragmatic approach since there are
phase space regions where the lowest order systematic expansion
fails~\cite{Baglio:2017bfe} that is signaled by the cross section
being negative! Notwithstanding, the use of the quadratic
contributions of the dimension--six operators is justified if the new
physics is strongly interacting; see for instance
Refs.~\cite{Biekoetter:2014jwa, Contino:2016jqw, Falkowski:2016cxu}.
Indeed, this result is consequence of Naive Dimensional
Analysis~\cite{Manohar:1983md} and some simple power counting
analysis. \smallskip In any case, Ref.~\cite{Brehmer:2015rna} shows
that the analysis of the LHC data in terms of dimension--six operators
is adequate in almost all realistic weakly coupled scenarios, except
in the high energy tails of distributions.  \smallskip

\begin{figure}[h!]
\centering
\includegraphics[width=0.9\textwidth]{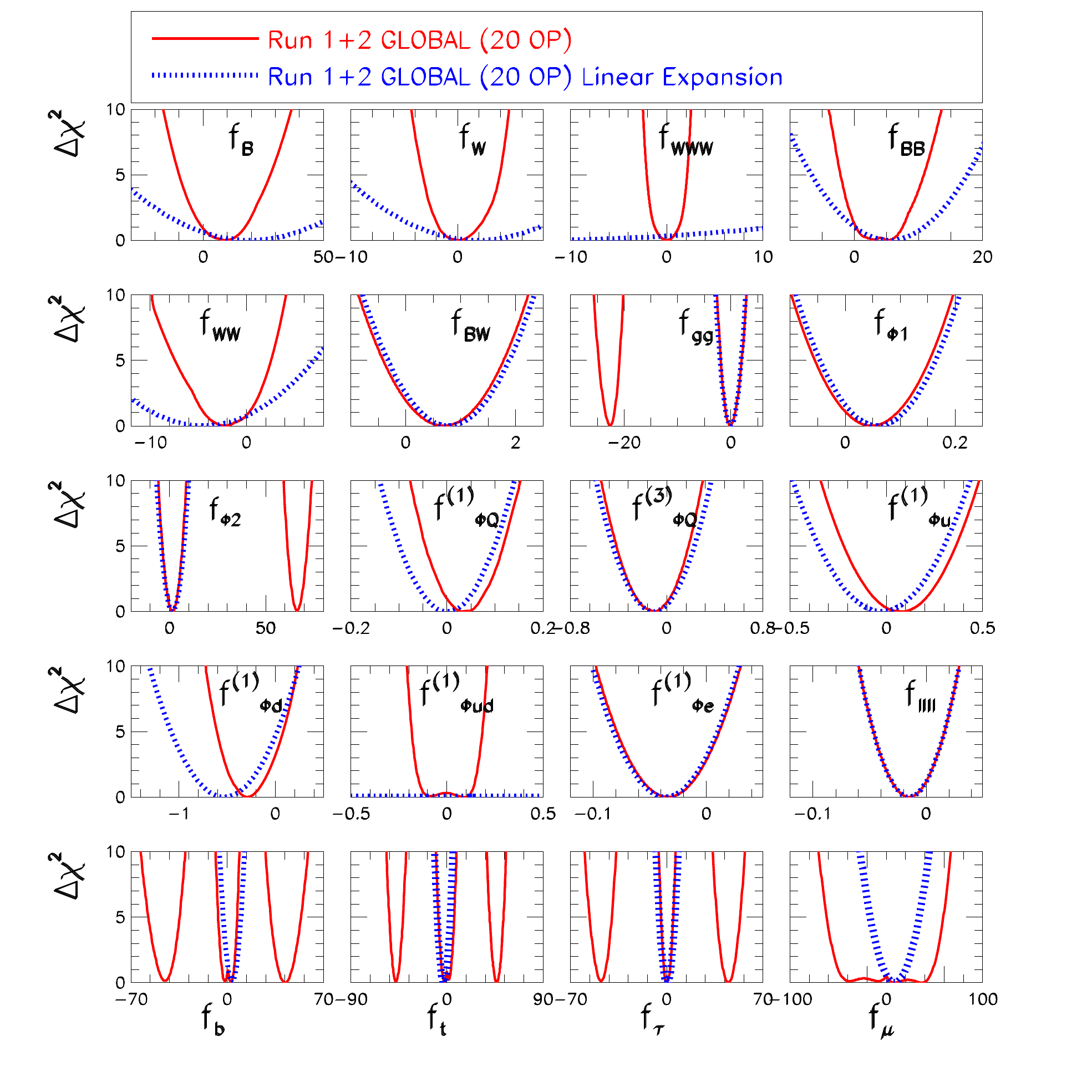}
\caption{$\Delta \chi^2$ dependence on the bosonic Wilson coefficients
$f/\Lambda^2$  as indicated in each panel. 
The results using only the linear terms in the anomalous couplings are
indicated by the blue dotted curves. The red solid
curves stands for the fits keeping the quadratic contributions of the
anomalous couplings to the observables. The full data set was used in
both cases.}
  \label{fig:linear}
\end{figure}

At this point we would like to understand the importance of keeping
the anomalous quadratic terms in the evaluation of the observables.
The result of this exercise certainly depends on the amount of data
available.  To this end, we redid our twenty-parameter fit using only
the contributions to the observables at linear order on the Wilson
coefficients; our results are depicted in
Fig.~\ref{fig:linear}. Comparing the results of the dashed curves in
Fig.~\ref{fig:linear} with the green curves in Fig.~\ref{fig:fer_r2}
we can see that the $\Delta \chi^2$ distributions as a function of
$(\fbw,\fpone,\fqthree,\fqone,\fur,\fdr,\fer,\fllll)$ are very much
the same as obtained using only the EWPD (see also
Table~\ref{tab:ranges}). This shows that the contributions to these
parameters due to diboson and Higgs data arise mainly from the
quadratic terms. Moreover, within the input range of variation of the
parameters in the analysis, the operator ${\cal O}_{\Phi,ud}$ is not
bounded if the observables are evaluated using just the linear terms
of their Wilson coefficients while and ${\cal O}_{WWW}$ is only very
weakly constrained. This happens because the dominant contributions of
these operators are to helicity amplitudes to which the SM does not
contribute~\cite{Alves:2018nof}.  \smallskip

The results in Fig.~\ref{fig:linear} show that the Higgs data is
already precise enough to determine $f_{\Phi,2}/\Lambda^2$ and
$f_{GG}/\Lambda^2$ in the linear approximation with the quadratic
terms being subdominant. Moreover, as expected, at linear order there
are no degenerate solutions in these couplings and the allowed regions
encompass only the SM. Same applies to the Yukawa-like operators so
the corresponding coefficients $f_f/\Lambda^2$ have no degenerate
solutions.  And therefore the $f_\mu/\Lambda^2$ seems better
determined. \smallskip

We also find that at present the Higgs data is precise enough to bound
the couplings $f_{BB}/\Lambda^2$ and $f_{WW}/\Lambda^2$ using the
linear evaluation of the observables, however, the size of the 95\% CL
allowed area increases by a factor of $\simeq 2$.  Similarly the
bounds on $f_B/\Lambda^2$ and $f_W/\Lambda^2$ are a factor of
${\cal O}(3-4)$ weaker than those obtained when we include the
quadratic terms~\cite{Corbett:2013pja}. Nevertheless, this is already
very interesting since the diboson production alone does only lead to
very mild bounds on these last two couplings when not including the
quadratic contributions, which indicates again the relevance of the
Higgs observables.\smallskip

\section{Summary}
\label{sec:summary}

In this work we have performed a comprehensive analysis of the observables
related to the electroweak sector, which at present allows for
precision tests of the couplings between electroweak gauge bosons and
fermions, triple electroweak gauge couplings and the couplings of the
Higgs to fermions and gauge bosons. This includes low energy
electroweak precision measurements as well as LHC data on gauge boson
pair production and Higgs observables.  In total, the GLOBAL analysis
of EWPD and EWDBD and Higgs results from LHC Run 1 encompasses 64
observables and including Run 1 and 2, 122 observables. \smallskip

\begin{table}
\begin{tabular}{|c|c|c|c|c|c|}\hline
  Operator & \multicolumn{5}{c|}{95\% CL $f/\Lambda^2$ (TeV$^{-2}$)}\\
\hline
    & EWPD & EWPD+EWDBD & GLOBAL Run 1 &
  GLOBAL Run 1+2 & GLOBAL Run1+2 Linear \\\hline
   $\ollll$ & (-0.043, 0.013) & (-0.043, 0.013) & (-0.043, 0.013)
  & (-0.043, 0.013) & (-0.043, 0.013)\\
   $\oer$ &  (-0.075,0.011) & (-0.075,0.007)& (-0.077,0.009)
  &  (-0.075,0.007) &  (-0.077,0.005)  \\
  $\opone$ & (-0.040,0.15)  &(-0.040,0.15) & (-0.043,0.15)& (-0.044,0.14)
  & (-0.034,0.15)  \\ 
  $\obw$  &(-0.32,1.7) &(-0.27,1.7)  & (-0.32,1.7) & (-0.30,1.7)
  & (-0.21,1.8)  \\
  $\oqthree$ & (-0.60,0.12)& (-0.45,0.13)  &(-0.49,0.11) &(-0.38,0.15)
  &(-0.41,0.19)\\
  $\oqone$ & (-0.083,0.10) &(-0.034,0.11) & (-0.049,0.11)&(-0.036,0.11)
  &(-0.089,0.088)\\ 
  $\odr$ & (-1.2,-0.13) &(-0.64,-0.007)  & (-0.73,0.02) & (-0.56,0.04)
  & (-1.0,-0.03)\\
  $\our$ &(-0.25,0.37) &(-0.17,0.37)  & (-0.22,0.38) & (-0.19,0.33)&
 (-0.32,0.25)\\
  $\oud$ & --- &(-0.17,0.17)  & (-0.29,0.29)  & (-0.18,0.18)
  &        ---      \\
   $\ob$ & --- & (-7.8,34)   & (-12,34)& (-8.3,26) & (-31,70)\\
   $\ow$ & --- & (-3.9,3.5)   & (-5.2,3.5)&(-3.0,3.7)& (-9.5,13) \\ 
   $\owww$ & ---& (-1.9,2.0)  & (-2.6,2.5)&(-1.9,2.0)&(-64,36)\\
   $\obb$ & --- & --- &(-2.5,13)&(-1.7,10)& (-5.4,16)\\
   $\oww$ & --- & --- &(-10,3.7)&(-6.7,2.1)&(-15,5.8)\\
   $\ogg$ & --- & --- &(-25,-17)$\,\cup\,$(-4.7,2.1) &(-25,-21)$\,\cup\,$(-1.7.1,8)&(-1.8,1.7)\\ 
   $\optwo$ & --- & --- &(-1.1,10) $\,\cup\,$(55,72)& (-3.2,6.2)$\,\cup\,$(62,71) &(-3.7,6.9)\\
  $\obo$ & --- & --- & (-62,-20)$\,\cup\,$(-12,11)$\,\cup\,$(23,45) &
  (-56,-36)$\,\cup\,$(-6.1,6.7)$\,\cup\,$(33,52) &(-2.2, 9.2)\\
  $\ot$ & --- & --- & (-64,-35)$\,\cup\,$(-19,20)$\,\cup\,$(37,59)&
  (-53,-42)$\,\cup\,$[-7.4,6.2]$\,\cup\,$(40,52) &(-8.3,2.4)\\  
  $\ota$ & --- & --- & (-59,-31) $\,\cup\,$(-5.8,7.8)$\,\cup\,$(32,50)&
  (-55,-41)$\,\cup\,$(-3.7,4.3)$\,\cup\,$(37,52) &(-4.8,5.4)\\ 
   $\omu$ & --- & --- &  ---  & (-50,57) &(-14,31)\\\hline   
\end{tabular}
\caption{95\% allowed ranges for the Wilson coefficients for the
  different analysis performed in this work. For  $\ot$
  we show in the 5th column the three discrete ranges allowed
  when no contribution of $tH$
  is included in the ATLAS cross section ratio.
  Including the $tH$ contribution under the assumptions discuss in the text
  selects the range around zero which we mark with square brackets.}
\label{tab:ranges}
\end{table}

We work in the framework of effective lagrangians with a linear
realization of the electroweak symmetry. At dimension six, and assuming
that the new operators do not introduce new tree level sources of
flavor violation nor violation of universality of the weak current,
the global analysis involves the 20 operators in Eq.~\eqref{eq:leff}
of which 8 contribute to EWPD (Eq.(\eqref{eq:leff-ewpd}), 4 additional
enter in the combination with the LHC EWDBD (Eq.(\eqref{eq:leff-tgc}),
and the 20 operators enter once the Higgs observables are
considered. \smallskip

\begin{figure}[h!]
\centering
\includegraphics[width=0.9\textwidth,height=0.33\textheight]{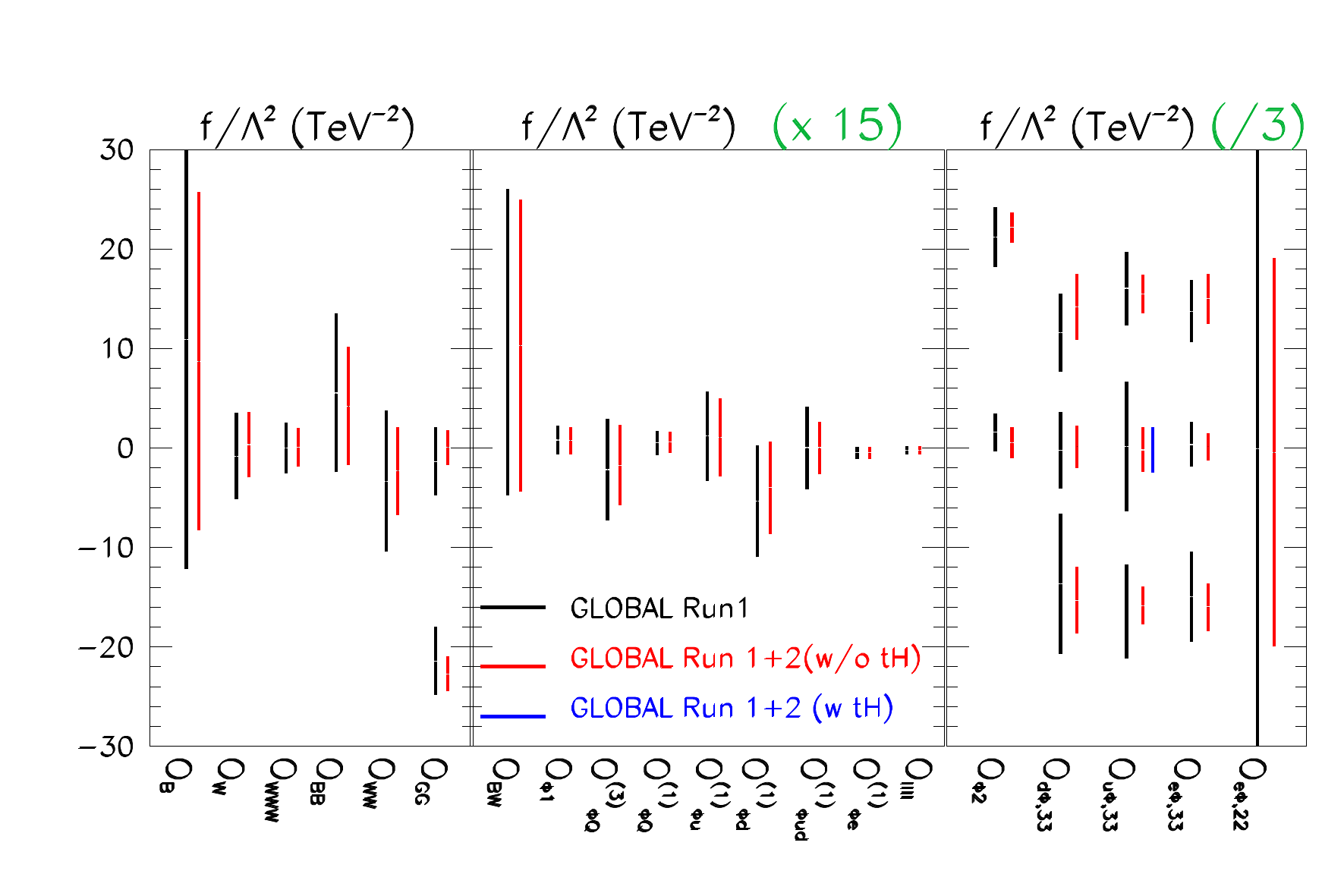}
\caption{Allowed 95\% CL ranges for each of the Wilson coefficients from the
20 parameter global analysis. We show results from the global EWPD+LHC
analysis including Run 1 EWDB and Higgs observables only (black bars)
and including both Run 1 and 2 EWDB and Higgs observables (red line).
For the top Yukawa-like operator $\ot$ we show as a blue bar the allowed
range when considering the contribution from $tHj$ and $tHW$ in the
to the top-Higgs associate production. To help graphical display the
ranges in the central (right) panels are multiplied (divide) by a factor
30 (3) as indicated on the top.}
\label{fig:ranges}
\end{figure}

Altogether the analyses show no statistically significant source of
tension with the SM. We find for the SM a $\chi^2_{\rm SM}=118$ (71.2)
for the 122 (64) observables in the GLOBAL analysis of EWPD and EWDBD
and Higgs results from LHC Run 1+2 (only Run 1).  Including the 20
Wilson coefficients in the fit, we find $\chi^2_{{\cal L}_{eff}}=91$
(52.3).  As a consequence, bounds on the Wilson coefficients can be
imposed.  The 95\% allowed ranges for the 20 Wilson coefficients
(profiled from the global 9-,12-, or 20-dimensional likelihoods) are
listed in Table~\ref{tab:ranges}. The corresponding allowed 95\% CL
ranges for the global analysis with Run 1 and Run 1 + 2 EWDBD and
Higgs observables are also graphically displayed in
Fig.~\ref{fig:ranges}. \smallskip

In brief, we find:
\begin{itemize}

\item The coefficients of the eight operators contributing to the EWPD
  are those better determined, as could be anticipated. The inclusion
  of the LHC EWDBD and Higgs observables has negligible impact on
  those operators contributing to the couplings of leptons, $\ollll$,
  $\oer$, $\opone$, and $\obw$.

\item The impact of the inclusion of the LHC results is still minor
  but not negligible for the EWPD bounded operators involving gauge
  couplings to quarks. In particular for $f^{(1)}_{\Phi d}/\Lambda^2$,
  under the assumption of generation universality a non-zero value for
  this coefficient is favored in the EWPD analysis coming from the
  2.7$\sigma$ discrepancy between the observed $A_{\rm FB}^{0,b}$ and
  the SM.  On the contrary, LHC observables involving this operator,
  are fully consistent with the SM what results in the shift and
  reduction of its globally allowed range.

\item The operator $\oud$ induces right-handed charged current
  couplings for quarks and it can only be bound via its quadratic
  contributions. Including those in the LHC observables its Wilson
  coefficient can be bounded with precision comparable to that of the
  other operators affecting gauge-quark couplings.

\item We have performed a novel analysis of the Run 2 ATLAS
  36.1 $\rm fb^{-1}$ data on
  transverse mass distribution of $W^+W^-$ and $W^\pm Z$ in the
  leptonic channel ~\cite{ATLAS:2018ogj, Aaboud:2017gsl} which allows
  for further tests of the TGC's.  The results are shown in
  Fig.~\ref{fig:tgv_r2}.  We find that the combined ATLAS Run 2
  diboson data constrains the operator coefficients with precision
  similar (a bit better indeed) to that of the full Run 1 analysis.
\item Inclusion of the Run 2 results in the global analysis
  results into a reduction of the allowed range for the coefficients of the 
  bosonic operators $\ob$, $\ow$, $\owww$, $\obb$, and $\oww$ by
  20--30\%.

\item The allowed values for $\fgg/\Lambda^2$ and $\fptwo/\Lambda^2$
  present each two discrete ranges originated by the degeneracy (it is
  a quasi-degeneracy for $\fptwo$) associated with the reverse of the sign
  of the $Hgg$ (Eq.\eqref{eq:vert-gluglu})  and $HXX$ ($X=f,V$)
  (Eqs.\eqref{eq:vert-hww},\eqref{eq:vert-yuk}), respectively.  Barring
  that degenerate solutions these are the best determined coefficients
  for operators not contributing to EWPD.

\item The allowed values for the coefficients for the Yukawa-like
  operators $\obo$ and $\ota$ ($\fbo$, and $\fta$), have a two folded
  degeneracy associated
  with the reverse of the sign of the corresponding $Hff$ coupling
  (see Eq.\eqref{eq:vert-yuk}) in combination with the reverse of the
  sign of the $HVV$ coupling. This results into the three discrete
  allowed ranges in Table~\ref{tab:ranges}. For the $\omu$ coefficient,
  $\fm$, the data is not
  precise enough to resolve the three solutions.

\item For $\ot$, the inclusion of the incipient $tH$ data can break
  those degeneracies on $\ft$, this is, in the determination of the
  top-Higgs coupling.

\item The last column in Table~\ref{tab:ranges} shows the allowed
  ranges when only the terms linear in the Wilson coefficients are
  kept in the observables. For those operators constrained by EWPD the
  bounds are just that of the EWPD as at LHC they are mainly
  constrained by its quadratic contribution. For the operators without
  degenerate solutions the bounds become weaker but are still within
  the same order or magnitude. Exceptions are $\oud$ and $\owww$ which
  become very weakly bounded as their dominant contributions at LHC
  are to helicity amplitudes which do not interfere with the SM
  ones. Keeping only the linear contribution to the observables does
  not allow for the degenerated solutions associated with the sign
  flip of the Higgs couplings. Consequently $\fgg/\Lambda^2$,
  $\fptwo/\Lambda^2$, $\fb/\Lambda^2$, $\ft/\Lambda^2$,
  $\fta/\Lambda^2$, and $\fm/\Lambda^2$ appear to be better
  constrained.

\item Our results show the importance of the Higgs data to constrain
  the TGC operators $\ob$ and $\ow$ when the LHC observables are
  evaluated using only the linear terms in the anomalous
  couplings. This extends the previous results in
  Refs.~\cite{Corbett:2013pja,deCampos:1997ez}.
\end{itemize}  

Altogether we find that the increased integrated luminosity gathered at 13 TeV
allows us to obtain more stringent bounds on a larger set of anomalous
interactions and to perform new tests of the SM. We look
forward for the  release of the complete dataset accumulated at Run 2. 

\acknowledgments 

We thank Tyler Corbett for discussions. We are indebted to Juan Gonzalez-Fraile
for his help with comparison with previous analysis.
M.C.G-G and N.R.A thank the group at Universidade de Sao Paulo for
their hospitality.  O.J.P.E. wants to thank the group at the Universitat
de Barcelona for the hospitality during the final stages of this work.
This work is supported in part by Conselho Nacional de Desenvolvimento
Cient\'{\i}fico e Tecnol\'ogico (CNPq) and by Funda\c{c}\~ao de Amparo
\`a Pesquisa do Estado de S\~ao Paulo (FAPESP) grants 2012/10095-7 and
2017/06109-5, by USA-NSF grant PHY-1620628, by EU Networks FP10 ITN
ELUSIVES (H2020-MSCA-ITN-2015-674896) and INVISIBLES-PLUS
(H2020-MSCA-RISE-2015-690575), by MINECO grant FPA2016-76005-C2-1-P
and by Maria de Maetzu program grant MDM-2014-0367 of ICCUB.
A. Alves thanks Conselho Nacional de Desenvolvimento
Cient\'{\i}fico (CNPq) for its financial support, grant 307265/2017-0.


\bibliography{references}
\end{document}